\documentstyle[psfig]{mn}

\newif\ifAMStwofonts

\def\arcs{$\prime\prime$}

\ifoldfss
  \ifCUPmtlplainloaded \else
    \NewTextAlphabet{textbfit} {cmbxti10} {}
    \NewTextAlphabet{textbfss} {cmssbx10} {}
    \NewMathAlphabet{mathbfit} {cmbxti10} {} 
    \NewMathAlphabet{mathbfss} {cmssbx10} {} 
  \fi
  \ifAMStwofonts
    \ifCUPmtlplainloaded \else
      \NewSymbolFont{upmath} {eurm10}
      \NewSymbolFont{AMSa} {msam10}
      \NewMathSymbol{\upi}     {0}{upmath}{19}
      \NewMathSymbol{\umu}     {0}{upmath}{16}
      \NewMathSymbol{\upartial}{0}{upmath}{40}
      \NewMathSymbol{$\leq$slant}{3}{AMSa}{36}
      \NewMathSymbol{\geqslant}{3}{AMSa}{3E}
           \let\oldleq=$\leq$
           
      \let$\leq$=$\leq$slant \let\le=$\leq$slant
       
    \fi
  \fi
\fi 

\ifnfssone
  \newmathalphabet{\mathit}
  \addtoversion{normal}{\mathit}{cmr}{m}{it}
  \addtoversion{bold}{\mathit}{cmr}{bx}{it}
  \newmathalphabet{\mathbfit} 
  \addtoversion{normal}{\mathbfit}{cmr}{bx}{it}
  \addtoversion{bold}{\mathbfit}{cmr}{bx}{it}
  \newmathalphabet{\mathbfss} 
  \addtoversion{normal}{\mathbfss}{cmss}{bx}{n}
  \addtoversion{bold}{\mathbfss}{cmss}{bx}{n}
  \ifAMStwofonts
    \ifCUPmtlplainloaded \else
      \UseAMStwoboldmath
      \makeatletter
      \new@mathgroup\upmath@group
      \define@mathgroup\mv@normal\upmath@group{eur}{m}{n}
      \define@mathgroup\mv@bold\upmath@group{eur}{b}{n}
      \edef\UPM{\hexnumber\upmath@group}
      \new@mathgroup\amsa@group
      \define@mathgroup\mv@normal\amsa@group{msa}{m}{n}
      \define@mathgroup\mv@bold\amsa@group{msa}{m}{n}
      \edef\AMSa{\hexnumber\amsa@group}
      \makeatother
      \mathchardef\upi="0\UPM19
      \mathchardef\umu="0\UPM16
      \mathchardef\upartial="0\UPM40
      \mathchardef$\leq$slant="3\AMSa36
      \mathchardef\geqslant="3\AMSa3E
           \let\oldleq=$\leq$
           
      \let$\leq$=$\leq$slant \let\le=$\leq$slant

    \fi
  \fi
\fi 

\ifnfsstwo
  \DeclareMathAlphabet{\mathbfit}{OT1}{cmr}{bx}{it}
  \SetMathAlphabet\mathbfit{bold}{OT1}{cmr}{bx}{it}
  \DeclareMathAlphabet{\mathbfss}{OT1}{cmss}{bx}{n}
  \SetMathAlphabet\mathbfss{bold}{OT1}{cmss}{bx}{n}
  \ifAMStwofonts
    \ifCUPmtlplainloaded \else
      \DeclareSymbolFont{UPM}{U}{eur}{m}{n}
      \SetSymbolFont{UPM}{bold}{U}{eur}{b}{n}
      \DeclareSymbolFont{AMSa}{U}{msa}{m}{n}
      \DeclareMathSymbol{\upi}{0}{UPM}{"19}
      \DeclareMathSymbol{\umu}{0}{UPM}{"16}
      \DeclareMathSymbol{\upartial}{0}{UPM}{"40}
      \DeclareMathSymbol{$\leq$slant}{3}{AMSa}{"36}
      \DeclareMathSymbol{\geqslant}{3}{AMSa}{"3E}
           \let\oldleq=$\leq$
           
      \let$\leq$=$\leq$slant \let\le=$\leq$slant

    \fi
  \fi
\fi 

\ifCUPmtlplainloaded \else
  \ifAMStwofonts \else 
    \def\upi{\pi}
    \def\umu{\mu}
    \def\upartial{\partial}
  \fi
\fi

\title[V, R, I and H$\alpha$ photometry]
  {V, R, I and H$\alpha$ photometry of circumnuclear star forming regions
 in four galaxies with different levels of nuclear activity.}
\author[Angeles I. D\'{\i}az et al.]
  {Angeles I. D\'{\i}az,$^1$
  Mar \'{A}lvarez \'{A}lvarez,$^1$ Elena Terlevich,$^2$ Roberto Terlevich,$^{3 ,}$\thanks{Visiting Professor, INAOE, Puebla, M\'exico}
  \newauthor 
  Miguel S\'{a}nchez Portal,$^{1,4}$ Itziar Aretxaga$^2$\\
  $^1$Dpto. de F\'{\i}sica Te\'{o}rica, C-XI, Universidad Aut\'{o}noma de Madrid, Cantoblanco, 28049-Madrid, Spain\\
  $^2$ INAOE, Tonantzintla, Ap. Postal 51 y 216, Puebla, M\'exico.\\
  $^3$ Institute of Astronomy, Madingley Road, CB3 OHA Cambridge, U.K.\\
  $^4$ LAEFF/INTA, P.O. Box 50727, 28080 Madrid, Spain.}
\date{Accepted . Received 1999}
\pagerange{\pageref{firstpage}--\pageref{lastpage}}
\pubyear{1994}

\def\LaTeX{L\kern-.36em\raise.3ex\hbox{a}\kern-.15em
    T\kern-.1667em\lower.7ex\hbox{E}\kern-.125emX}

\begin{document}

\label{firstpage}

\maketitle

\begin{abstract}

 We present photometry, in V,R,I continuum bands and in the H$\alpha$ + [NII]
emission lines, for a sample of four circumnuclear star forming regions (CNSFR), located in galaxies with different kinds of activity in their nuclei: NGC~7469 
(Sy 1), NGC~1068 (Sy 2), NGC~7177 (LINER) and  NGC~3310 (Starburst).

H$\alpha$ luminosities for the CNSFR range from 0.02 to 7 $\times$ 10$^{40}$ ergs s$^{-1}$ (uncorrected for internal extinction), comparable to those observed in other galaxies, with NGC~7177 showing the lowest luminosity in average.
No systematic differences in the broad band colours are found for the CNSFR in the different galaxies, except for those in NGC~3310 which 
are considerably bluer. This fact is found to be partially due to a younger 
stellar population.

The colours have been analysed in the light of theoretical evolutionary 
synthesis models.  In some cases they can be reproduced by single populations   with ages between 7 and 300 Myr and modest values of extinction (0.5-1.5 mag). However, in many cases, this population is unable to provide the observed  equivalent widths of H$\alpha$, which require the presence of a younger population.

In the case of  NGC~1068, NGC~7177, and NGC~7469, acceptable fits are found for a two-burst population model at solar metallicity: the younger burst with an age between 2 and 8 Myrs, provides the bulk of the ionization and the older one (8 - 20 Myrs) is responsible for the continuum light at wavelengths longer than H$\beta$. 
The age difference between both populations is around 5-7 Myr and the younger burst involves from 3 to 61\% of the total mass of the cluster. This would be consistent with the younger burst being originated by the supernova
activity from the previous one.
Models of this kind also reproduce the regions in NGC~3310, but for younger 
ionizing population ages (between 1 and 3 Myr) and a metallicity 0.25 times 
solar.

In most cases an excess in the observed  (R-I) colour over the model predicted 
one is found, which is not consistent with a normal reddening law. If this 
excess is atributted to the red supergiants present in the older population, 
this seems to imply that this population is not properly taken into account by the models.

 In this two-population scenario there seems to be a trend for the 
circumnuclear star forming regions of NGC~3310 (starburst), NGC~1068 
(Seyfert 2) and NGC~7469 and NGC~7177 (Seyfert 1 and LINER respectively) to 
be progressively older. Whether this implies a relation between the evolutionary state of the regions and the nuclear type of the parent galaxy remains to be explored.

\end{abstract}

\begin{keywords}
 Galaxies: circumnuclear regions -- Active galactic nuclei -- HII regions.
\end{keywords}

\section{Introduction}

Early in 1967, S\'{e}rsic and Pastoriza realised that the inner parts of some
spiral galaxies showed higher star formation rate than usual, and that this 
star formation was
frequently arranged in a ring or pseudo-ring pattern. This fact seems to 
correlate with the presence of bars in early type spiral galaxies \cite{comb,atha}. 
Computational models that simulate the behaviour of gas in galactic
potentials show that nuclear rings may appear as a consequence of
matter infall, due to resonances present at the bar edges. In these 
places the gas loses angular momentum and falls inwards.
If, in addition, two inner Lindbald resonances exist -- which usually happens
in early spirals -- gas will condense between both instead of falling
directly into the nucleus; this higher gas density may lead to a
higher star formation rate, through interacting molecular clouds \cite{comb}
or gravitational fragmentation of the ring \cite{elme}.
If no inner Lindbald resonance is present, or a very massive central object
dominates the dynamics, gas is allowed to continue falling further 
into the potential well, and eventually may give rise to a nuclear, instead of 
circumnuclear, starburst or feed an active galactic nucleus \cite{tele,elm}.
In this scenario, the most puzzling problem is explaining the presence
of nuclear and circumnuclear activity at the same time  since,
in order to give rise to a circumnuclear ring, the gas needs to stop 
their infalling in an inner Lindbald resonance and, to generate a
nuclear burst or active nucleus, it must continue getting inwards.

%
%

\begin{table*}
 \begin{minipage}{100mm}
 \caption{Characteristics of the sample}
 \label{symbols}
 \begin{tabular}{@{}ccccccc}
  \bf Galaxy & \bf Morphological & \bf pc/$^{\prime\prime}$ & \bf Nuclear &\bf B$_T$ & 
  \bf D (Mpc) & \bf E(B-V)$_{\rm gal}$ \\
           &   \bf Type     &       &  \bf Type   &       &                     \\
 \hline
  NGC 1068 &(R)SA(rs)b    & 87.8   & Sy2       & 9.61  & 18.1$^{1}$ & 0.05 \\
  NGC 3310 & SAB(r)bc pec & 60.6   & Starburst & 11.15 & 12.5$^{2}$ & 0.00\\
  NGC 7177 & SAB(r)b      & 74.2   & LINER     & 12.01 & 15.3$^{3}$ & 0.23\\
  NGC 7469 & (R')SAB(rs)a & 315.12 & Sy1       & 13.00 & 65.0$^{4}$ & 0.12\\
 \hline
 \end{tabular}
  Notes to table 1:\\
1.- Sandage \& Tammann 1975 \\
2.- Bottinelli et al. 1984 \\
3,4.- NED \footnote{The National Extragalactic Database (NED), is operated by the Jet Propulsion Laboratory, California Institute of Technology, under contract with the National Aeronautics and Space Administration.} Ho = 75  Km s$^{-1}$ Mpc.
 \end{minipage}
 \end{table*}

	However, nuclear activity and the presence of circumnuclear structures undergoing star formation, are simultaneous phenomena in a large number of
galaxies \cite{arse} and their mutual influence and the role played by each of them in the
dynamics, and/or the energy balance of the galaxy, are still open questions.
Three possibilities have been put forward: 
	1) The nuclear activity has its origin in the supernova remnants 
product of the intense star formation rate experienced by the circumnuclear
regions of the galaxy \cite{weed,norm}.
	2) The star formation in the circumnuclear regions has been induced 
by gas and radiation ejected by the active nucleus.
	3) Both phenomena are not related to each other.

A possible way to elucidate this matter may be through the study of the general 
properties and evolutionary state of CNSFR in galaxies with different levels of 
nuclear activity. Specifically, their age and metallicity can provide important 
clues about the sequence of appearance of both phenomena: nuclear and circumnuclear activity.
 
A few studies have been made comparing circumnuclear/``hot spot" and disc 
HII region populations ( e.g. Kennicutt et al. 1989; Mayya 1994) and several studies of circumnuclear rings have been 
made, using observations obtained from the ground and the Hubble Space Telescope (HST) , to try to characterize 
the stellar content of every single burst \cite{bart,buta,holt,gonz,gar}; however, not a systematic comparison  between the CNSFRs of different emission-line type galaxies has so far been made.

For this study we have  selected 4 nearby galaxies with confirmed circumnuclear 
star forming rings and different nuclear type: Seyfert 1 (NGC~7469), Seyfert 2 
(NGC~1068), LINER (NGC~7177) and Starburst (NGC~3310) to perform a comparative 
study.

NGC~7469 is a well studied Seyfert 1 galaxy.
Maps of [OIII] and  H$\alpha$ show a ring outlined by star forming regions,
with a radius of about 1-2 Kpc that surrounds the nucleus \cite{maud}.This ring seems to be related to a bar observed in the K band (2.1 $\mu$m) \cite{wil}. Very 
Large Array (VLA) observations  show this ring to split into individual radio emitting knots \cite{wils}. 

NGC~1068 is one of the best studied Seyfert galaxies. In 1985  Antonucci and Miller detected broad permitted lines in its  polarized spectrum, and since then it has become the representative example of the unified model for Active Galactic Nuclei (AGN). Interestingly enough, a strong  
CaII triplet in absorption has also been observed and atributted to the presence of red supergiant stars \cite{tdt90}.
It shows, as well, one of the most prominent circumnuclear rings, with a radius of about 1 Kpc together with an infrared (IR) bar \cite{scov,plan}.
The fact that all these features occur simultaneously in this object makes of 
it an excellent laboratory for the study of possible interrelations between them.

No many bibliographical notes can be found on NGC~7177, but for 7 nuclear
star forming regions observed by Hodge (1982). It also appears in some
compilations  \cite{vert,pogg}.

Finally, NGC~3310 is a good example of an overall low metallicity galaxy, with a high rate of star 
formation and very blue colours. It is also a prominent source of
X Rays and Ultraviolet radiation.
The circumnuclear region was studied spectrophotometrically by Pastoriza et al. (1993).
The brightest HII region was reported by Terlevich {\it et al.} (1990) to show the IR CaII 
triplet in absorption. No presence of nuclear
bars has been reported till now and therefore the merging with a low metallicity
object about 18 Myrs ago, is proposed as the fuelling mechanism
for the nuclear burst (Smith {\it et al.} 1996).

 In this paper we present photometric results through broad band V,R,I and narrow H$\alpha$ filters on a total of 68 regions, distributed on the four galaxy rings.
Observations and reduction details are described in Section 2. The results are presented in Section 3 and discussed in Section 4.  Finally, the summary and main conclusions of the work  are presented in Section 5.

\section{Observations and reductions}

 The observations were made as part of two observing runs in 1988 and 1990
using a blue sensitive GEC CCD at the f/15 Cassegrain focus
of the 1.0 m Jacobus Kaptein Telescope (JKT) of the Isaac Newton Group at the
 Observatorio del Roque de los Muchachos, La Palma. The CCD had
578 $\times$ 385 pixels 22 $\mu$m wide. The scale 
obtained with this instrumental configuration is 0.3 arcsec pixel$^{-1}$,
and the CCD field is 2.89$^{\prime}$ $\times$ 1.92$^{\prime}$. The main
characteristics of the sample galaxies are summarized in Table 1. Column 1 
gives the galaxy identification; column 2 the morphological type as listed 
in the 3$^{rd}$ Reference Catalog of Bright Galaxies (RC3; de Vaucouleurs {\it et al.} 1991); column 3 the linear size in parsecs per arc second;
column 4 the nuclear type; column 5 the total B magnitude (RC3) and column 6, the distance taken from the 
references given in the table. Finally, column 7 gives the adopted galactic 
extinction towards the galaxy (see this section below).

Observations and reductions are described in detail in S\'anchez Portal et al. (1999) and are outlined in what follows.

 All observations were made under photometric conditions, and the seeing was estimated using stars present in each frame.
 Each night photometric \cite{land} and spectrophotometric \cite{mass,ston}
standards were observed to perform the corresponding calibrations. We also took dome and sky flat-field images as well as zero exposure time frames to set the
bias level.

  Images in the broad V, R, and I and narrow  H$\alpha$ and continuum filters
were obtained for every galaxy, except in the case of NGC~3310, for which no
H$\alpha$ continuum image was taken. A journal of observations is presented in table 2. The filter characeristics are given in Table 3. Columns 1 to 4 give, 
respectively, the filter name, central wavelength and full width at half maximum
(FWHM) in angstroms, and maximum transmission. The H$\alpha$ filter 
includes the lines of [NII] at $\lambda\lambda$ 6548,6584 \AA\ although the 
latter enters the filter at about half transmission. The typical value of the 
[NII] $\lambda$ 6548 \AA\ line in HII regions is at most 1/2 of H$\alpha$. 

The data reduction was carried out using  IRAF and MIDAS routines following the 
standard steps: subtraction of the bias level, flat-field division and sky 
background removal which was performed by averaging the mean count values in 
several boxes in the outer parts of each frame. Later on, cosmic rays were 
removed, the point spread function (PSF) was estimated using stellar images in each frame, and flux calibration was performed.

Both atmospheric and galactic extinction corrections were applied. The first 
one using the 
extinction coefficients provided by La Palma Observatory, and the second one 
using the values of E(B-V) taken from Burstein and Heiles (1984; see Table 1) and the 
galactic extinction curve of Seaton (1979). No internal absorption correction 
was attempted.

%
%

\begin{table}
\begin{minipage}{100mm}
 \caption{Journal of observations}
 \label{symbols}
 \begin{tabular}{@{}cccc}
  \bf Galaxy & \bf Filter & \bf exp. time (s) & \bf seeing (\arcs)\\ 
 \hline
 \bf NGC 1068 & V    & 100   & 2.1      \\
 \            & R    & 100   & 2.1      \\
 \            & I    & 100   & 2.1      \\
 \            & LINE & 800   & 1.2       \\
 \            & CONT & 800   & 1.2      \\
 \bf NGC 3310 & V    & 200   & 1.02     \\
 \            & R    & 300   & 1.02      \\
 \            & I    & 300   & 1.02     \\
 \            & LINE & 700   & 1.02      \\
 \bf NGC 7177 & V    & 600   & 1.2\\
 \            & R    & 600   & 1.2      \\
 \            & I    & 600   & 1.0      \\
 \            & LINE & 2000  & 1.3       \\
 \            & CONT & 2000  & 1.4      \\
 \bf NGC 7469 & V    & 500   & 1.3 \\
 \            & R    & 300   & 1.3      \\
 \            & I    & 300   & 1.3      \\
 \            & LINE & 800   & 1.5       \\
 \            & CONT & 1000  & 1.5      \\
\hline
\hline
 \end{tabular}
\end{minipage}
 \end{table}


\section{Results}

\subsection{HII region photometry}

H$\alpha$ frames were continuum subtracted in order to obtain a net H$\alpha$ 
line image. H$\alpha$ continuum images were available for NGC~1068, NGC~7177 
and NGC~7469. In the case of NGC~3310, a continuum frame was constructed from 
the broad band R image as explained in Terlevich {\it et al.} (1991). In all 
cases the two images to be subtracted were previously aligned according to the 
offsets derived from the centroids of a bidimensional gaussian fit to the field 
star images.

Figure 1 shows the H$\alpha$ line contour maps for the four observed galaxies. 
Individual HII regions are marked and labelled. In the images, North is to 
the top and East is to the left.

In the four studied galaxies the CNSFRs are arranged in a ring pattern. The 
biggest ring, with a mean radius of 2 Kpc, is found in NGC~7469, followed by 
that of NGC~1068, with a mean radius of 1.8 Kpc. Smaller rings are found for 
NGC~7177 and NGC~3310, with mean values of about 0.8 Kpc. The rings of 
NGC~7469, NGC~1068 and NGC~3310 are almost circular, while that of NGC~7177 
shows a more distorted morphology.

Star forming regions have been identified on  H$\alpha$ line images and their 
fluxes have been measured.
When computing the sizes and fluxes of these regions, three 
main problems were encountered.\\
1. Photometry software packages are not very efficient to measure
fluxes from diffuse objects in a non-constant background.\\
2. In very crowded regions, deciding the limits of two adjacent regions
is almost arbitrary.\\
3. Even in isolated regions, the determination of their radii is always difficult.

%
%

\begin{table}
\begin{minipage}{100mm}
 \caption{Filter characteristics}
 \label{symbols}
 \begin{tabular}{@{}cccc}
 \bf Filter & \bf $\lambda_{\rm c}$(\AA) & \bf FWHM (\AA) & \bf $\tau_{\rm max}$ (\%)\\ 
 \hline
 V                                    &  5470  &  938 & 80 \\
 R                                    &  6455  & 1253 & 87 \\
 I                                    &  8300  & 1813 & 85 \\
 H$\alpha$ (v$_{\rm r}$ = 0 Km/s)     &  6563  &   53 & 50 \\
 H$\alpha$ (v$_{\rm r}$ = 4000 Km/s)  &  6652  &   49 & 54 \\
 \hline
 \end{tabular}
\end{minipage}
 \end{table}


	Theoretically, a radiation bound HII region, can be characterized 
by its Str\"{o}mgren radius. But, in practice, they show a bright
core, surrounded by a more extended and dimmer region. The border
radius between them can only be distinguished in
isolated and not distorted regions. This can be seen in Figure 2 where we have
plotted flux (in arbitrary units) {\it vs} radius for HII regions of 
different morphologies.

Taking this into account, we have followed different criteria in order to 
determine the HII region sizes and fluxes. In isolated regions, we have computed the  H$\alpha$ flux inside circular apertures of different radii \footnote[1]{Due to software limitations no circular
regions could be integrated interactively, so we used square boxes instead. Radius stands
for the equivalent radius of the circular region having the same area as the square one.
$r = \sqrt{Area/\pi}$}. 

In the plot of  flux {\it vs}
radius ( Fig 2) , an asymptotic behaviour can be seen. For most observed 
regions the asymptotic radius is reached when the flux falls down to 10 \% 
of the flux at the central pixel. We have taken this radius as the radius of 
the region, and the flux inside it as the region flux.

When the morphology of the region is distorted (see NGC~7469 \#2 in 
Fig 2), we have added all the flux inside a circular aperture up to the 
isophote corresponding to 10\% of the central value, even though the minimum could be lower.
When two regions are very close to each other we have computed their flux together if the minimum signal along the line that
joins the two central pixels is higher than 50\% of the smaller 
maximun. In all other cases, we have integrated both regions separately.

Once the sizes and H$\alpha$ fluxes of every region were measured, 
H$\alpha$ equivalent widths -- computed by division of the H$\alpha$ line frame by the corresponding continuum image -- were also obtained.
As already mentioned above, no H$\alpha$ continuum image was available for 
NGC~3310. In this case the H$\alpha$ equivalent width was calculated using 
the R filter as described in Bessell (1990). The two methods yield slightly 
different results. This can be seen in Figure 3 where we plot the equivalent 
width of H$\alpha$ derived using its corresponding continuum, EW(cont), against that calculated using the R-band image as continuum, EW(R). The solid line stands for EW(cont)=EW(R), while the best linear 
regression fit gives $ EW(cont) = 0.77 EW(R) + 12.3 $ \AA . Outlying regions 6 and 22 in 
NGC~1068 present an anomalous behaviour. Both are faint and are located in 
the direction of the ionization cone subtended by the nuclear radiation. They 
have not been taken into account in computing the fit and have been excluded 
from the general study.

In order to measure the broad band fluxes, all the frames 
were carefully aligned as explained above, and the corresponding fluxes were 
measured inside the apertures defined on the H$\alpha$ line frame.  	
Subsequently, the broad band magnitudes were computed and these magnitudes 
were subtracted to get the colours.

Sizes and distances from the galactic nucleus 
\begin{footnote}{ The galactic nucleus has been defined as 
the maximum of the H$\alpha$ flux distribution}
\end{footnote}
(in arc seconds), V magnitudes, broad-band colours, H$\alpha$ fluxes (in erg s$^{-1}$ cm$^{-2}$), and H$\alpha$ equivalent widths (in \AA )  
computed
in the described  way are listed in Table 4 for every detected HII region in each 
observed galaxy. 

We have estimated the background contribution to the computed magnitudes
from the observed intensity radial profiles (S\'anchez-Portal {\it et al.} 1999) for each galaxy. The average contribution is about 10\% and the  maximum contribution has been found for region \# 2 in NGC~7177 amounting to about 
40 \%  in the I band. This region is faint and located close to the bulge of the galaxy. The mean background colours for NGC~7177 are V-R= 0.4 and R-I=0.65 
(S\'anchez-Portal et al. 1999), therefore the background subtracted colours of 
the region are about 0.03 magnitudes redder than computed without taking this 
contribution explicitely into account.

In fact, the main source of error in this work comes from the determination of the radius
for each HII region. Therefore, in order to estimate realistic errors, 
the fluxes inside the adopted radius plus and minus one pixel  
have also been computed for each region.  We have adopted
as errors the differences between the two calculated extreme values. These 
errors are also listed in Table 4. This procedure effectively takes into account the contribution by any local variable background.

R frames were further corrected for the contribution of H$\alpha$ +[NII] emission 
using the H$\alpha$ line image as explained in Terlevich {\it et al.} (1991).
In most cases this correction amounts to only 0.1 mag.

\subsection{Characteristics of the CNSFRs}

Angular sizes and galactocentric radii, apparent magnitudes and H$\alpha$ fluxes have been converted to 
linear quantities, absolute magnitudes and H$\alpha$ luminosities using the 
distance to each galaxy given in Table 1, and are listed in Table 5. The 
quoted linear size corresponds to the aperture radius used for the photometry.

Linear diameters range from 82.4 to 684.8 pc. The size of the largest region in 
each galaxy decreases from 684.8 pc in NGC~1068 to 182 pc in NGC~3310, with 
NGC~7469 and NGC~7177 showing intermediate values (428.6 pc and 252.2 pc 
respectively).

H$\alpha$ luminosities range from 2 $\times$10$^{38}$ to 7$\times$ 10$^{40}$ erg s$^{-1}$. 
These values are similar to those found for other CNSFRs in galaxies 
({\it e.g.} Gonz\'alez-Delgado {\it et al.} 1995). Most of them have 
logL(H$\alpha$) $>$ 39.0, which places them into the category of supergiant 
HII regions as defined by Kennicutt (1983). As can be seen in Fig 4, for three of the galaxies:
NGC~1068, NGC~3310 and NGC~7469, the peak H$\alpha$ luminosity is found at 
logL(H$\alpha$) = 39.5, while for NGC~7177 the luminosities seem to be, in
average, lower by an order of magnitude.
The cumulative luminosity functions for the CNSFRs in each of the galaxies  have been fitted by laws of the form 
$ N(L) \propto  L^{1-\alpha}$. The values of $\alpha$ obtained from the fits
are close to the value of 2 found by Kennicutt (1983), except in the case of 
NGC~7469 for which $\alpha$ = 2.5. An even steeper slope is found by 
Gonz\'alez Delgado {\it et al.} (1997) for the high luminosity disc HII regions 
of this galaxy. However, given the small number of points entering the fits and 
the systematic errors inherent to the method, this fact might not be significant.

The H$\alpha$ luminosity {\it vs} size relation can be seen in Fig 5. For each 
of the galaxies the data can be fitted in the log-log plane by a regression 
line of slope between 2 and 2.5. However, better fits are obtained for the 
regions in NGC~1068 and NGC~3310 (a=2.2, logL$_o$(H$\alpha$) =35.6) and 
NGC~7177 and NGC~7469 (a=2.2, logL$_o$(H$\alpha$) = 34.5)  taken separately.
This seems to indicate that, for a given H$\alpha$ luminosity, the CNSFRs in 
NGC~7177 and NGC~7469 are larger and more diffuse than in the other two 
galaxies in which the regions seem to be more compact.

Regarding the properties of the stellar continuum, some information can be 
obtained by looking at the I {\it vs} R-I colour-magnitude diagram, since 
our R images are corrected for H$\alpha$ + [NII] emission and the I frame is 
almost emission-line free. Figure 6 shows that in this diagram the regions of 
NGC~3310 are well separated from the rest, showing lower continuum luminosity 
and bluer colours. This can be partially due to their low metallicity (about  
1/10 solar; Pastoriza {\it et al.} 1993).
The CNSFRs in the other three galaxies show R-I colours between 0.46 and 0.88.
Since these colours are 
in most cases redder than the average value for their galactic discs -- 0.50, 
0.62 and 0.65 for NGC~1068, NGC~7469 and NGC~7177 respectively --, they cannnot be attributed to a substantial background contribution not taken into account.
Actually, any further background subtraction would produce even redder colours. 

H$\alpha$ equivalent widths also provide information about the stellar 
continuum properties. By combining the four histograms at the left of Figure 4, it can be seen that the distribution of EW(H$\alpha$) is double peaked. While NGC~1068 and NGC~3310 show values centered 
about logEW(H$\alpha$) = 2.4, those corresponding to NGC~7177 and NGC~7469  are 
clustered around logEW(H$\alpha$)= 1.5 implying that either their regions are
more evolved or the contribution by a non ionizing population is more important for the regions in these last  two galaxies.

In fact, when plotting logEW(H$\alpha$) {\it vs} V-I colour 
(Fig 7) a relation seems to 
emerge with regions with larger equivalent widths showing bluer colours. Since 
the equivalent width of H$\alpha$ is a good age indicator for ionized regions, 
we are tempted to attribute this relation to age rather than  metallicity.

\subsection{Comparison with earlier photometry}

There is not much broad band photometry of HII regions. Our results can be 
compared with those of Mayya (1994) and Telles \& Terlevich (1997). The objects
studied by Mayya are mainly disc HII regions for which V, R and H$\alpha$ 
photometry is given. On the other hand, Telles \& Terlevich provide V, R, I and 
H$\alpha$ photometry for HII galaxies.

Figure 8 shows the absolute V magnitude (left panel) and the logarithm of the 
equivalent width of H$\alpha$ (right panel) {\it vs} V-R colour for our objects 
together with those of Mayya and Telles \& Terlevich. HII 
galaxies are the brightest objects in the plot, while some disc HII regions are 
at the lower luminosity end. CNSFRs show an intermediate behaviour. Except for 
some low luminosity disc HII regions which look redder, all the objects show similar V-R colours. These regions also have rather large H$\alpha$ equivalent
widths and so their excess R flux might be due to the contribution of the 
H$\alpha$ emission.

Regarding equivalent widths, the regions of NGC~3310 have values comparable to 
those found in HII galaxies while the CNSFRs in the rest of the galaxies have 
EW(H$\alpha$) similar to those found in disc HII regions.

\section{Discussion}

\subsection{Evolutionary synthesis models}

Our results can be interpreted with the help of evolutionary population 
synthesis models. Leitherer \& Heckman (1995) provide broad-band colours and 
H$\alpha$ luminosities and equivalent widths for different populations that 
can be directly compared with our data. Models for both instantaneous bursts 
and continuous star formation computed with different Initial Mass Functions 
(IMFs) are given. The assumed metallicities range from 0.1 to 2 times solar.
We have used for comparison the models with metallicity 0.25 solar for the 
regions of NGC~3310 in agreement with oxygen abundance determinations for some 
of their CNSFRs (Pastoriza {\it et al.} 1993). Solar metallicity models have been adopted for the rest of the galaxies. In all cases a Salpeter  IMF has been assumed.  

Figure 9 (top left panel) shows the run of instantaneous burst subsolar (solid line) and solar (dashed
line) models in the V-R {\it vs} R-I colour-colour diagram as compared with 
data. Each tick on the lines corresponds to 1 Myr from 1 to 10 Myr and 5 Myr 
for 10 to 40 Myr The end of the line corresponds to an age of about 300 Myr.
It can be seen that, as the cluster evolves from 1 to 3 Myr, its V-R 
colour decreases at a constant R-I value, as the V flux increases. From then onwards the general trend is for the colours to get redder with time. 
The reddest colours reached by the subsolar model are about V-R=0.25 and 
R-I=0.5. For the solar metallicity models the evolutionary track moves vertically upwards for ages between  6 and 7 Myr and downwards for ages 
between 10 and 15 Myr. 
 
 The effect of 1 magnitude extiction would move the model points up and to the right by the amount indicated by the arrow in the plot.  
This diagram is also affected by the emission of [OIII] in the V filter. This line is significant only at low metallicities and very young ages \cite{stas} 
but its maximum effect would be to move downwards the points by about 0.005 mag in V-R. This uncertainty is in most cases smaller than the average observational error also shown in the plot.

In the absence of reddening, the colours of the regions of NGC~3310 can be reproduced by single burst populations of age between 7 and 10 Myr. Different 
amounts of extinction, between 0.5 and  1.5 mag, would allow younger ages.
The extinction derived for some of these regions from spectroscopic 
data are between 0.4 and 1.15 mag (Pastoriza {\it et al.} 1993).

For the rest of the regions, the observed colours can be adequately reproduced by single burst populations  with age between 7 to 300 Myrs, the maximum age computed by Leitherer and Heckman, if different amounts of extinction are 
assumed.

 Continuous star formation models provide, in general, a poorer fit to the data.
Older ages ( 30 to 100 Myr) are found using this kind of model  
and visual extinctions larger than 2 mag are, in many cases,  necessary to reproduce the observed colours.

Further insight can be gained by looking at the behaviour of equivalent
width of 
H$\alpha$ {\it vs} R-I colour predicted by the models (Fig. 9, top right panel), since it provides information about the ionizing stars. The large equivalent 
widths observed in the regions of NGC~3310 can only be reproduced by very young 
populations with ages between 3 and 5 Myr. Values of extinction less than 0.7 mag are therefore implied. The ionizing population of the regions in NGC~1068, NGC~7469 and NGC~7177 seems to be progressively older, being in most cases between 6 and 9 Myr. The models with contiuous star formation which best 
reproduce the observed colours cannot reproduce the moderate values of the  equivalent widths of H$\alpha$ which 
are  observed and therefore have not been pursued further.

On the other hand, as already mentioned in the previous section, a 
relation seems to exist between logEW(H$\alpha$) and R-I. If interpreted 
as an extinction effect, this would imply an increasing extinction from 
NGC~1068 to NGC~7177. However, no relation is apparent between logEW(H$\alpha$) 
and V-R (Fig. 8) where the effects of extinction should be more important. 
The difference in R-I  can also be attributed 
to an increasing contribution of  red supergiant stars (Garc\'\i a-Vargas 
{\it et al.} 1998) which could be associated with an underlying non-ionizing 
population. 

\subsection{The case for a two-burst population model}

The two graphs just discussed generate a three dimensional space in which two dimensions (the R-I and V-R colours) represent basically the stellar continuum properties, while the third one (equivalent width of H$\alpha$) stands for the importance of the gas emission  relative to the stellar continuum.
For a model with given metallicity and initial mass function, a zero-age single burst population is represented in this space by a point. As time goes on, 
this point should move along the track given by the theoretical model. 
Several effects could move data away from this theoretical line: the presence
of reddening,  the contribution of emission lines to the continuum band filters 
and the contribution of an extra non-ionizing stellar continuum flux. The 
second of these effects is actually negligible, except for very low metallicities and very young ages. 

Taking this into account, consistent solutions for single burst populations can be found only for some of the regions of NGC~1068. In this case the burst ages are 
between 6 and 7 Myr and the derived visual extinctions are always less than 
1 mag. For the rest of the regions, no consistent solutions for single burst
populations are found, since different amounts of extiction are needed to 
reconcile models and data in each of the diagrams.

Therefore we have computed simple two-population models by combining  young 
ionizing populations, with ages between 1 and 10 Myr, and non-ionizing underlying populations, with ages between 8 and 20 Myr (for ages older that 20 Myr the V-R and R-I colours remain practically constant within our observational
errors), taken in different proportions. 
For each population pair, we have calculated the value of the equivalent width of H$\alpha$ with increasing contribution of the young component to the total mass. For each studied region, the intersection of the line of observed EW(H$\alpha$) with the model lines gives the possible solutions.
The procedure is illustrated in Fig. 11.
Once the possible models have been selected, we have calculated the corresponding V-R and R-I colours and have compared them with the observed ones. Then we have chosen the solution than minimizes the difference between observed and calculated colours. No reddening is assumed for any of the two populations.

The selected models for each region and their characteristics are listed in Table 6, and shown in Fig. 10. The best models for the regions in NGC~3310 correspond to the combination of a young population 2.5 Myr old and an older one 8 Myr old, taken in different proportions as indicated in the table. For NGC~1068 the best combination corresponds to bursts of 5 and 9 Myr and these ages are increased to 8 and 15 Myr for NGC~7177 and NGC~7469. In most cases, the age difference between the two components may be consistent with the time elapsed in the models for the first supernovae to explode. According to Leitherer \& Heckman's models this time is around 3.6 Myr both for metallicities
solar and 0.25 solar and the expected supernova rate is 10$^{-3}$ yr$^{-1}$ 
during a period of about 25 Myr.

 Columns 3 and 4 in Table 6 give the model computed colours while columns 5 and 6 give the difference between observed and computed colours.
In most cases the computed V-R colour agrees with the observed one within the errors, this is also true for the R-I colour of the regions of NGC~3310. However, for the rest of the regions a considerable excess on the observed R-I colour over the computed one is found, which is not consistent with a normal reddening law. (R-I) colours of red supergiant stars are between 0.4 and 
1.00 (D\'\i az {\it et al.} 1989), which is actually the range covered by our observed 
HII regions. Therefore, if this near infrared excess is atributted to the presence of a 
red supergiant population, this seems to imply that this population is not 
properly taken into account by the models.

Two burst populations have also been found for the circumnuclear regions of 
NGC~3310 studied spectroscopically by Pastoriza {\it et al.} (1993); also 
 Mayya 
\& Prabhu (1996), from a study of disc HII regions, find in most cases 
evindence 
for an accompanying population rich in red supergiants from a previous burst and Kennicutt et al. (1989) mention the possibility that star formation in ``hot spots" takes place
over timescales much longer than in normal disc HII regions in order to explain the lower values of H$\alpha$ equivalent widths found for the former objects.

If this two-population  scenario is adopted, there is some indication for the CNSFR in 
NGC~7177 and NGC~7469,  LINER and  Seyfert type 1 respectivey, to be older 
than those of NGC~1068, Seyfert type 2, which in turn look older than those 
of NGC~3310, a starburst galaxy. More observations are obviously needed in 
order to test if the history of the circumnuclear star formation activity is 
related to the nuclear type of the parent galaxy.

\section{Summary and conclusions}

We have studied 68 star forming complexes around four galactic nuclei showing
different degrees of activity. For all of them we have obtained H$\alpha$ fluxes, luminosities and equivalent widths, sizes, V, R, and I magnitudes and V-R and R-I colours.

The linear diameters of the regions range from 82.4 to 684.8 pc and their
H$\alpha$ luminosities range from 2$\times$ 10$^{38}$ to 7$\times$ 10$^{40}$ ergs, which places them in the category of supergiant HII regions according to the classification 
given by Kennicutt (1983).

The regions of NGC~7177 are about an order of magnitude less luminous than the others and it seems that, for a given H$\alpha$ luminosity, the regions in NGC~7177 and NGC~7469 are larger and more diffuse than in the other two galaxies in which the regions seem to be more compact.
The distribution of logEW(H$\alpha$) in the observed CNSFRs is double-peaked: the regions in NGC~1068 and NGC~3310 show values centered at about 2.4 while those of NGC~7177 and NCG~7469 are clustered around 1.5 implying than either these last regions are more evolved or the contribution by a non ionizing population is more important.

Regarding the  colours it seems that the regions in 
NGC~3310 show bluer continuum colours than the rest. This is probably due to a mixed effect of a younger age and a lower metallicity.

If we compare the observed colours (V-R and R-I) and  H$\alpha$ equivalent widths with those
predicted by theoretical evolutionary synthesis models  we find that, except 
for a few  regions in NGC~1068, in most of the observed regions the data
cannot be adequately reproduced by a combination of a single burst of star formation and a normal reddening law.
 The best fitting models involve a composite population with two clusters of ages 5 and 9 Myr in the 
case of NGC~1068, 2.5 and 8 Myr for NGC~3310, and 8 and 15 Myr in the cases of NGC~7177 and NGC~7469.
The contribution to the total cluster mass by the younger component is found 
to be between 3 and 61 \%. 
 The age difference between the two assumed bursts is consistent with the time elapsed in the models until the  supernovae explosions from the first burst take place.

Under this two-population scenario the CNSFR ofNGC~7469 and NGC~7177 (Seyfert 1 and LINER respectively) are found to be 
older than the corresponding ones in  NGC~1068 (Seyfert 2) and NGC~3310 
(starburst). More observations are needed to establish if the age of the CNSFR and the nuclear type of the galaxy are related to one another.

\section*{Acknowledgements}

The JKT is operated in the island of La Palma by the Issac Newton Group in the Spanish Observatorio del 
Roque de los Muchachos of the Instituto de Astrof\'\i sica de Canarias. We 
would like to thank CAT for awarding observing time.
We also would like to thank VILSPA and LAEFF for kind support with the first reduction of our data.  We also thank an anonymous referee for her/his comments 
and suggestions. 

E.T. is grateful to an IBERDROLA Visiting Professorship 
to UAM during which this work was completed.
This work has been partially supported by DGICYT project PB-96-052.

%
%

\begin{table*}
 \begin{minipage}{150mm}
 \caption{Observed sizes, magnitudes and H$\alpha$ fluxes of measured HII regions}
 \label{symbols}
 \begin{tabular}{@{}cccccccc}
 \hline
  \bf region&\bf radius&\bf distance from &\bf V&\bf V-R&\bf R-I&\bf F(H$\alpha$)  &\bf EW(H$\alpha$)  \\
  \bf number&\bf (\arcs)  &\bf centre (\arcs)    & & \bf (Mag.) & & \bf (erg $s^{-1} cm^{-2}$)& \bf (\AA) \\
\hline
&&&&NGC~1068&&&\\
\hline
\hline
1&1.7&(-17.4,-2.8)&16.5$\pm$0.4&0.47$\pm$0.02&0.65$\pm$0.001&2.2e-13&242.5\\
2&1.7&(-23.1,-2.4)&17.2$\pm$0.5&0.37$\pm$0.02&0.58$\pm$0.002&1.1e-13&180.6\\
3&1.7&(-20.4, 0.6)&16.6$\pm$0.4&0.40$\pm$0.001&0.72$\pm$0.01&1.7e-13&222.7\\
4&1.4&(-18.3, 3.9)&16.4$\pm$0.4&0.49$\pm$0.02&0.64$\pm$0.004&2.1e-13&199.4\\
5&3.9&(-19.5,11.1)&15.0$\pm$0.2&0.41$\pm$0.001&0.46$\pm$0.02&1.6e-12&369.9\\
6&--&--&--&--&--&--&--\\
7&1.5&(-10.8,-0.9)&15.2$\pm$0.4&0.37$\pm$0.008&0.59$\pm$0.01&2.6e-13&95.6\\
8&1.5&(-10.2, 4.2)&15.6$\pm$0.5&0.44$\pm$0.008&0.58$\pm$0.002&3.2e-13&135.9\\
9&1.5&(-11.7, 6.9)&15.2$\pm$0.5&0.38$\pm$0.03&0.60$\pm$0.002&3.1e-13&133.0\\
10&2.2&(-10.8,9.9)&15.0$\pm$0.4&0.38$\pm$0.008&0.62$\pm$0.003&6.1e-13&75.8 \\
11&1.6&(-7.5,12.3)&15.8$\pm$0.5&0.46$\pm$0.01&0.61$\pm$0.02&3.3e-13&153.4\\
12&0.9&(-5.4,12.6)&16.4$\pm$0.6&0.48$\pm$0.01&0.64$\pm$0.002&1.0e-13&147.6\\
13&2.0&(-3, 13.8)&16.0$\pm$0.5&0.46$\pm$0.0006&0.62$\pm$0.01&3.5e-13&142.5 \\
14&1.3&(5.4,9.3)&15.3$\pm$0.4&0.27$\pm$0.02&0.52$\pm$0.001&2.1e-13&107.3\\
15&1.3&(7.8,9.6)&15.6$\pm$0.5&0.36$\pm$0.002&0.49$\pm$0.01&2.2e-13&139.9   \\
16&1.3&(8.7,5.1)&15.7$\pm$0.6&0.29$\pm$0.01&0.52$\pm$0.007&1.7e-13&77.2   \\
17&1.3&(11.1,7.5)&16.4$\pm$0.6&0.35$\pm$0.02&0.52$\pm$0.009&1.6e-13&127.2 \\
18&1.3&(13.2,7.5)&16.8$\pm$0.6&0.45$\pm$0.002&0.46$\pm$0.008&1.6e-13&187.5 \\
19&2.4&(13.5,-7.8)&15.1$\pm$0.4&0.56$\pm$0.01&0.53$\pm$0.008&6.9e-13&163.9\\
20&2.0&(10.5,-11.7)&14.7$\pm$0.3&0.44$\pm$0.01&0.54$\pm$0.003&5.3e-13&135.6\\
21&1.6&(7.2,-12.9)&16.2$\pm$0.5&0.56$\pm$0.0004&0.53$\pm$0.02&2.5e-13&157.9 \\
22&--&--&--&--&--&--&--\\
23&1.3&(-5.1,-7.2)&16.3$\pm$0.6&0.45$\pm$0.002&0.62$\pm$0.0006&1.9e-13&126.2 \\
A&1.3&(13.5,-15.3)&17.0$\pm$0.7&0.45$\pm$0.002&0.49$\pm$0.008&9.9e-14&122.5  \\
B&1.6&(12.0,17.4)&16.6$\pm$0.5&0.45$\pm$0.02&0.50$\pm$0.003&1.6e-13&163.4   \\
\hline
&&&&NGC~3310&&&\\
\hline
\hline
1&0.7&(-10.5,2.7)&19.3$\pm$0.9&0.32$\pm$0.002&0.12$\pm$0.03&6.8e-14&953.4 \\
2&1.0&(-9.6,4.2)&17.6$\pm$0.5&0.37$\pm$0.005&0.13$\pm$0.008&1.5e-13&448.7\\
3&1.4&(-6.0,1.5)&16.3$\pm$0.4&0.26$\pm$0.03&0.15$\pm$0.03&5.4e-13&527.2   \\
4&1.0&(-3.0,4.8)&16.5$\pm$0.5&0.21$\pm$0.04&0.21$\pm$0.01&3.3e-13&426.5   \\
5&0.7&(-2.1,5.4)&17.2$\pm$0.9&0.23$\pm$0.009&0.23$\pm$0.04&1.8e-13&416.4   \\
6&1.0&(-0.3,6.3)&16.9$\pm$0.6&0.26$\pm$0.01&0.21$\pm$0.002&1.3e-13&229.3   \\
7&1.0&(3.3,6.3)&16.9$\pm$0.6&0.25$\pm$0.004&0.20$\pm$0.02&1.2e-13&229.8 \\
8&0.7&(3.0,1.5)&17.5$\pm$0.9&0.33$\pm$0.005&0.29$\pm$0.004&5.2e-14&167.9   \\
9&1.0&(4.5,0.0)&16.9$\pm$0.6&0.27$\pm$0.03&0.24$\pm$0.01&9.3e-14&178.3   \\
10&1.0&(2.4,-4.5)&16.9$\pm$0.7&0.34$\pm$0.01&0.25$\pm$0.007&9.3e-14&164.8 \\
11&1.0&(1.2,-6.6)&16.8$\pm$0.6&0.32$\pm$0.0004&0.24$\pm$0.0009&1.1e-13&180.0\\
12&1.4&(-2.4,-8.1)&16.0$\pm$0.4&0.30$\pm$0.6&0.14$\pm$0.6&2.9e-13&232.4\\
13&1.0&(-8.4,-6.0)&16.6$\pm$0.6&0.25$\pm$0.009&0.15$\pm$0.005&1.3e-13&191.8   \\
14&1.0&(-5.4,-1.5)&16.8$\pm$0.6&0.36$\pm$0.01&0.21$\pm$0.009&1.8e-13&257.5 \\
15&0.7&(-9.3,-1.5)&18.5$\pm$0.9&0.45$\pm$0.01&0.16$\pm$0.02&7.7e-14&489.6 \\
16&1.0&(-7.2,-0.3)&17.1$\pm$0.6&0.35$\pm$0.0006&0.13$\pm$0.02&3.3e-13&564.8 \\
17&0.7&(10.2,-0.9)&18.4$\pm$0.9&0.29$\pm$0.01&0.18$\pm$0.03&4.4e-14&328.8 \\
18&1.0&(8.4,-5.1)&16.9$\pm$0.6&0.25$\pm$0.008&0.09$\pm$0.02&2.8e-13&469.5 \\
19&1.7&(8.4,-7.5)&15.7$\pm$0.2&0.20$\pm$0.02&0.00$\pm$0.02&1.2e-12&626.8 \\
20&1.0&(-22.2,-)&17.6$\pm$0.5&0.34$\pm$0.004&0.09$\pm$0.04&1.5e-13&405.9 \\
21&1.0&(8.1,-13.2)&18.4$\pm$0.5&0.33$\pm$0.03&0.05$\pm$0.03&1.5e-13&788.4 \\
\hline
 \end{tabular}               
 \end{minipage}
 \end{table*}
 
\begin{table*}
\setcounter{table}{3}
 \begin{minipage}{150mm}
 \caption{Continued}
 \label{symbols}
 \begin{tabular}{@{}cccccccc}
 \hline
  \bf region&\bf radius&\bf distance from &\bf V&\bf V-R&\bf R-I&\bf F(H$\alpha$)  &\bf EW(H$\alpha$)  \\
  \bf number&\bf (\arcs)  &\bf centre (\arcs)    & & \bf (Mag.) & & \bf (erg $s^{-1} cm^{-2}$)& \bf (\AA) \\
\hline
&&&&NGC~7177&&&\\
\hline
\hline
1&1.4&(-17.1,2.1)&15.5$\pm$0.5&0.46$\pm$0.009&0.8$\pm$0.01&5.1e-14&25.8   \\
2&0.7&(-4.5,4.8)&17.6$\pm$0.9&0.44$\pm$0.03&0.69$\pm$0.008&7.9e-15&27.4   \\
3&1.0&(-3.0,6.0)&16.7$\pm$0.6&0.45$\pm$0.01&0.7$\pm$0.008&1.5e-14&23.6   \\
4&1.7&(-0.3,3.6)&15.0$\pm$0.4&0.44$\pm$0.002&0.78$\pm$0.006&6.7e-14&22.8   \\
5&1.7&(0.0,0.0)&15.0$\pm$0.4&0.50$\pm$0.02&0.88$\pm$0.009&7.3e-14&21.2   \\
6&1.4&(-6.0,-2.7)&17.0$\pm$0.5&0.43$\pm$0.006&0.71$\pm$0.006&1.8e-14&34.4   \\
7&1.7&(-2.7,-7.2)&16.5$\pm$0.4&0.37$\pm$0.006&0.71$\pm$0.004&3.6e-14&46.7   \\
8&1.4&(0.0,-8.7)&16.8$\pm$0.5&0.35$\pm$0.01&0.67$\pm$0.008&2.0e-14&33.1   \\
9&1.7&(5.7,-11.1)&15.3$\pm$0.1&0.35$\pm$0.005&0.56$\pm$0.009&2.1e-14&9.3   \\
10&1.7&(-13.8,-9.6)&17.7$\pm$0.4&0.29$\pm$0.006&0.56$\pm$0.02&3.5e-14&139.4\\
11&1.4&(16.5,8.1)&18.2$\pm$0.5&0.47$\pm$0.04&0.85$\pm$0.01&1.4e-14&79.7   \\
\hline
&&&&NGC~7469&&&\\
\hline
\hline
1&0.7&(2.1,3.9)&16.9$\pm$0.9&0.42$\pm$0.03&0.62$\pm$0.002&1.3e-14&53.0   \\
2&0.7&(0.0,4.2)&17.0$\pm$0.9&0.39$\pm$0.03&0.66$\pm$0.02&1.3e-14&63.3   \\
3&0.7&(-1.5,3.9)&17.1$\pm$1.0&0.42$\pm$0.01&0.62$\pm$0.02&5.8e-15&58.2   \\
4&0.7&(-2.7,3.6)&18.9$\pm$0.9&0.61$\pm$0.05&0.65$\pm$0.02&8.4e-15&50.1   \\
5&0.7&(-6.3,1.8)&19.1$\pm$0.9&0.43$\pm$0.01&0.58$\pm$0.02&5.0e-15&39.1   \\
6&0.3&(-8.7,-4.5)&19.9$\pm$2.0&0.31$\pm$0.03&0.48$\pm$0.05&2.0e-15&37.9   \\
7&0.3&(-3.6,-8.1)&20.6$\pm$2.0&0.41$\pm$0.09&0.6$\pm$0.1&1.5e-15&51.1   \\
8&0.7&(0.0,-7.5)&19.1$\pm$0.9&0.34$\pm$0.02&0.5$\pm$0.02&5.5e-15&47.6   \\
9&0.7&(4.2,-1.5)&18.5$\pm$0.9&0.50$\pm$0.02&0.59$\pm$0.02&8.4e-15&37.8   \\
10&0.5&(4.2,0.3)&17.5$\pm$0.7&0.54$\pm$0.03&0.60$\pm$0.01&2.6e-14&47.2   \\
11&0.7&(3.9,2.7)&18.6$\pm$1.0&0.59$\pm$0.05&0.71$\pm$0.003&8.8e-15&36.8   \\
\hline
 \end{tabular}               
 \end{minipage}
 \end{table*}
 

%
%

\begin{table*}
 \begin{minipage}{120mm}
 \caption{Linear sizes, absolute magnitudes and H$\alpha$ luminosities of 
the observed HII regions}
 \label{symbols}
 \begin{tabular}{@{}ccccccc}
 \hline
  \bf region&\bf radius&\bf distance&\bf M(V)&\bf M(R)&\bf M(I)&\bf logL(H$\alpha$)  \\
  \bf number&\bf (pc)  &\bf (pc)    & \bf (Mag.)& \bf (Mag.) & \bf (Mag.)& \bf (erg $s^{-1} cm^{-2}$) \\
\hline
&&&NGC~1068&&&\\
\hline
\hline
\    1   &149.3&1545.1   &-14.8&-15.3&-15.9&39.9\\
\    2   &149.3&2037.9   &-14.1&-14.4&-15.0&39.7\\
\    3   &149.3&1790.8   &-14.7&-15.1&-15.8&39.8\\
\    4   &149.3&1641.9   &-14.9&-15.4&-16.0&39.9 \\
\    5   &342.4&1968.9   &-16.3&-16.7&-17.2&40.8\\
\    6   &--&--&--&--&--&--\\
\    7   &131.7&951.0    &-16.1&-16.5&-17.1&40.0\\
\    8   &131.7&968.0    &-15.7&-16.1&-16.7&40.1\\
\    9   &131.7&1192.0   &-16.1&-16.4&-17.0&40.1\\
\    10  &193.2&1285.6   &-16.3&-16.6&-17.3&40.4\\
\    11  &140.5&1264.1   &-15.5&-16.0&-16.6&40.1\\
\    12  &$\leq$79.1&1202.9   &-14.9&-15.4&-16.0&39.6\\
\    13  &175.4&1239.2   &-15.2&-15.7&-16.3&40.1\\
\    14  &114.1&943.7    &-15.0&-16.3&-16.8&39.9\\
\    15  &114.1&1085.4   &-15.7&-16.0&-16.5&39.9\\
\    16  &114.1&884.9    &-15.5&-15.8&-16.4&39.8\\
\    17  &114.1&1175.5   &-14.9&-15.3&-15.8&39.8\\
\    18  &114.1&1332.2   &-14.5&-14.9&-15.4&39.9\\
\    19  &210.7&1368.1   &-16.2&-16.8&-17.3&40.4\\
\    20  &175.6&1379.5   &-16.6&-17.0&-17.6&40.3\\
\    21  &140.5&1296.3   &-15.1&-15.6&-16.1&40.0\\
\    22  &--&--&--&--&--&--\\
\    23  &114.1&774.2    &-15.0&-15.4&-16.0&39.9\\
\    A   &114.1&1790.5   &-14.3&-14.8&-15.2&39.6\\
\    B   &140.5&1854.7   &-14.7&-15.1&-15.6&39.8\\
 \hline
&&&NGC~3310&&&\\
\hline
\hline
\    1   &41.2&657.0   &-11.2&-11.5&-11.7&39.1   \\
\    2   &61.8&635.0   &-12.9&-13.3&-13.4&39.5   \\
\    3   &81.8&374.8   &-14.2&-14.4&-14.6&40.0   \\
\    4   &61.8&343.0   &-14.0&-14.2&-14.4&39.8   \\
\    5   &41.2&351.1   &-13.3&-13.5&-13.8&39.5   \\
\    6   &61.8&382.2   &-13.6&-13.9&-14.1&39.4   \\
\    7   &61.8&431.0   &-13.6&-13.8&-14.0&39.3   \\
\    8   &41.2&203.3   &-13.0&-13.3&-13.6&39.0   \\
\    9   &61.8&272.7   &-13.6&-13.9&-14.1&39.2   \\
\    10  &61.8&309.1   &-13.6&-13.9&-14.2&39.2   \\
\    11  &61.8&406.5   &-13.7&-14.0&-14.3&39.3   \\
\    12  &81.8&512.0   &-14.5&-14.8&-14.9&39.7   \\
\    13  &61.8&625.6   &-13.9&-14.1&-14.3&39.4   \\
\    14  &61.8&339.6   &-13.7&-14.1&-14.3&39.5   \\
\    15  &41.2&570.9   &-12.0&-12.4&-12.6&39.2   \\
\    16  &61.8&436.7   &-13.4&-13.8&-13.9&39.8   \\
\    17  &41.2&620.5   &-12.0&-12.3&-12.5&38.9   \\
\    18  &61.8&595.5   &-13.6&-13.9&-13.9&39.7   \\
\    19  &103.0&682.4  &-14.8&-15.0&-15.0&40.3   \\
\    20  &61.8&1480.5  &-12.9&-13.3&-13.2&39.4   \\
\    21  &61.8&938.5   &-12.1&-12.5&-12.5&39.4   \\
\hline
 \end{tabular}               
 \end{minipage}
 \end{table*}

\begin{table*}
\setcounter{table}{4}%
 \begin{minipage}{120mm}
 \caption{Continued}
 \label{symbols}
 \begin{tabular}{@{}ccccccc}
 \hline
   \bf region&\bf radius&\bf distance&\bf M(V)&\bf M(R)&\bf M(I)&\bf logL(H$\alpha$)  \\
  \bf number&\bf (pc)  &\bf (pc)    & \bf (Mag.)& \bf (Mag.) & \bf (Mag.)& \bf (erg $s^{-1} cm^{-2}$) \\
\hline
&&&NGC~7177&&&\\
\hline
\hline
\    1   &100.2    &1277.9   &-15.4&-15.9&-16.7&    39.2  \\
\    2   &$\leq$50.5&488.0   &-13.3&-13.8&-14.4&    38.3  \\
\    3   &75.7     &497.6    &-14.2&-14.7&-15.4&    38.6  \\
\    4   &126.1    &268.0    &-15.9&-16.4&-17.2&    39.3  \\
\    5   &126.1    &0        &-15.9&-16.4&-17.3&    39.3  \\
\    6   &100.2    &488.0    &-13.9&-14.4&-15.1&    38.7  \\
\    7   &126.1    &570.4    &-14.4&-14.8&-15.5&    39.0  \\
\    8   &100.2    &645.3    &-14.1&-14.5&-15.2&    38.8  \\
\    9   &126.1    &925.6    &-15.6&-16.0&-16.5&    38.8  \\
\    10  &126.1    &1246.9   &-13.2&-13.5&-14.0&    39.0  \\
\    11  &100.2    &1363.4   &-12.7&-13.2&-14.0&    38.6  \\
\hline
&&&NGC~7469&&&\\
\hline
\hline
    1   &214.3&1395.8   &-17.1&-17.5&-18.2& 39.8  \\
    2   &214.3&1323.5   &-17.1&-17.5&-18.1& 39.8  \\
    3   &214.3&1316.7   &-17.0&-17.4&-18.0& 39.5  \\
    4   &214.3&1418.0   &-15.2&-15.8&-16.4& 39.6  \\
    5   &214.3&2064.7   &-15.0&-15.4&-16.0& 39.4  \\
    6   &$\leq$107.1&3086.6   &-14.1&-14.4&-14.9& 39.0  \\
    7   &$\leq$107.1  &2793.2   &-13.5&-13.9&-14.5& 38.9  \\
    8   &214.3&2363.4   &-14.9&-15.3&-15.8& 39.4  \\
    9   &214.3&1405.4   &-15.5&-16.0&-16.6& 39.6  \\
    10  &214.3&1326.9   &-16.5&-17.1&-17.7& 40.1  \\
    11  &160.7&1494.8   &-15.5&-16.1&-16.8& 39.7  \\
\hline
 \end{tabular}               
 \end{minipage}
 \end{table*}


%
%

\begin{table*}
 \begin{minipage}{100mm}
 \caption{Mass contribution of the younger burst (M$_y$) to the total stellar mass of the HII region (M$_t$),computed colours, and differences (CD) between the observed colours and those given by the models.}
 \label{symbols}
 \begin{tabular}{@{}cccccc}
\bf Region & \bf My/Mt \%& \bf (V-R) & \bf (R-I) &\bf CD(V-R) &\bf CD(R-I)\\
\hline
&&NGC~1068&5/9 Myrs&&\\
\hline
1&61&0.40&0.33&0.07&0.32\\
2&46&0.44&0.37&-0.07&0.21\\
3&57&0.41&0.34&-0.01&0.38\\
4&51&0.43&0.36&0.06&0.28\\
5&92&0.31&0.26&0.1&0.20\\
6&--&--&--&--&--\\
7&22&0.50&0.41&-0.13&0.18\\
8&33&0.47&0.39&-0.03&0.19\\
9&33&0.47&0.39&-0.09&0.21\\
10&16&0.52&0.43&-0.08&0.19\\
11&38&0.46&0.38&0.00&0.23\\
12&37&0.46&0.39&0.02&0.25\\
13&35&0.47&0.39&-0.01&0.23\\
14&25&0.49&0.41&-0.22&0.11\\
15&35&0.47&0.39&-0.11&0.10\\
16&17&0.52&0.42&-0.23&0.10\\
17&31&0.48&0.40&-0.13&0.12\\
18&48&0.43&0.36&0.02&0.10\\
19&41&0.45&0.38&0.11&0.15\\
20&33&0.47&0.39&-0.03&0.15\\
21&40&0.46&0.38&0.10&0.15\\
22&--&--&--&--&--\\
23&31&0.48&0.40&-0.03&0.22\\
A&30&0.48&0.40&-0.03&0.09\\
B&41&0.45&0.38&0.00&0.12\\
\hline
&&NGC~3310&2.5/8 Myrs&&\\
\hline
1&52&0.24&0.09&0.08&0.03\\
2&26&0.29&0.15&0.08&-0.02\\
3&30&0.29&0.14&-0.03&0.01\\
4&24&0.30&0.15&-0.09&0.06\\
5&24&0.30&0.15&-0.07&0.08\\
6&13&0.32&0.17&-0.06&0.04\\
7&13&0.32&0.17&-0.07&0.03\\
8&9&0.33&0.18&0.00&0.11\\
9&9&0.33&0.18&-0.06&0.06\\
10&8&0.33&0.18&0.01&0.0  7\\
11&9&0.33&0.18&-0.01&0.06\\
12&13&0.32&0.17&-0.02&-0.03\\
13&10&0.33&0.18&-0.08&-0.03\\
14&14&0.33&0.17&0.03&0.04\\
15&28&0.29&0.15&0.16&0.01\\
16&32&0.28&0.14&0.07&-0.01\\
17&19&0.31&0.16&-0.02&0.02\\
18&27&0.29&0.15&-0.04&-0.06\\
19&35&0.27&0.13&-0.07&-0.13\\
20&23&0.30&0.16&0.04&-0.07\\
21&44&0.26&0.11&0.07&-0.06\\
\hline
 \end{tabular}               
 \end{minipage}
 \end{table*}

\begin{table*}
\setcounter{table}{5}
 \begin{minipage}{100mm}
 \caption{Continued}
 \label{symbols}
 \begin{tabular}{@{}cccccc}
\bf Region & \bf My/Mt \% & \bf (V-R) & \bf (R-I) &\bf CD(V-R) &\bf CD(R-I)\\
\hline
&&NGC~7177&8/15 Myrs&&\\
\hline
1&12&0.34&0.48&0.12&0.32\\
2&13&0.34&0.48&0.10&0.21\\
3&11&0.34&0.47&0.11&0.23\\
4&11&0.34&0.47&0.10&0.31\\
5&9&0.33&0.47&0.17&0.41\\
6&18&0.36&0.48&0.07&0.23\\
7&26&0.38&0.48&-0.01&0.23\\
8&17&0.35&0.48&0.00&0.19\\
9&3&0.31&0.47&0.04&0.09\\
10&56&0.41&0.48&-0.12&0.08\\
11&57&0.44&0.50&0.03&0.35\\
\hline
&&NGC~7469&8/15 Myrs&&\\
\hline
1&32&0.39&0.49&0.03&0.13\\
2&40&0.41&0.49&-0.02&0.17\\
3&36&0.40&0.49&0.02&0.13\\
4&29&0.39&0.49&0.22&0.16\\
5&21&0.37&0.48&0.06&0.1\\
6&20&0.36&0.48&-0.05&0.00\\
7&30&0.39&0.49&0.02&0.11\\
8&27&0.38&0.48&-0.04&0.02\\
9&20&0.36&0.48&0.14&0.11\\
10&27&0.38&0.48&0.26&0.12\\
11&19&0.36&0.48&0.23&0.23\\
\hline
 \end{tabular}               
 \end{minipage}
 \end{table*}


\newpage


%
%

\begin{figure*}
\begin{minipage}{170mm}
  \hbox{\psfig{figure=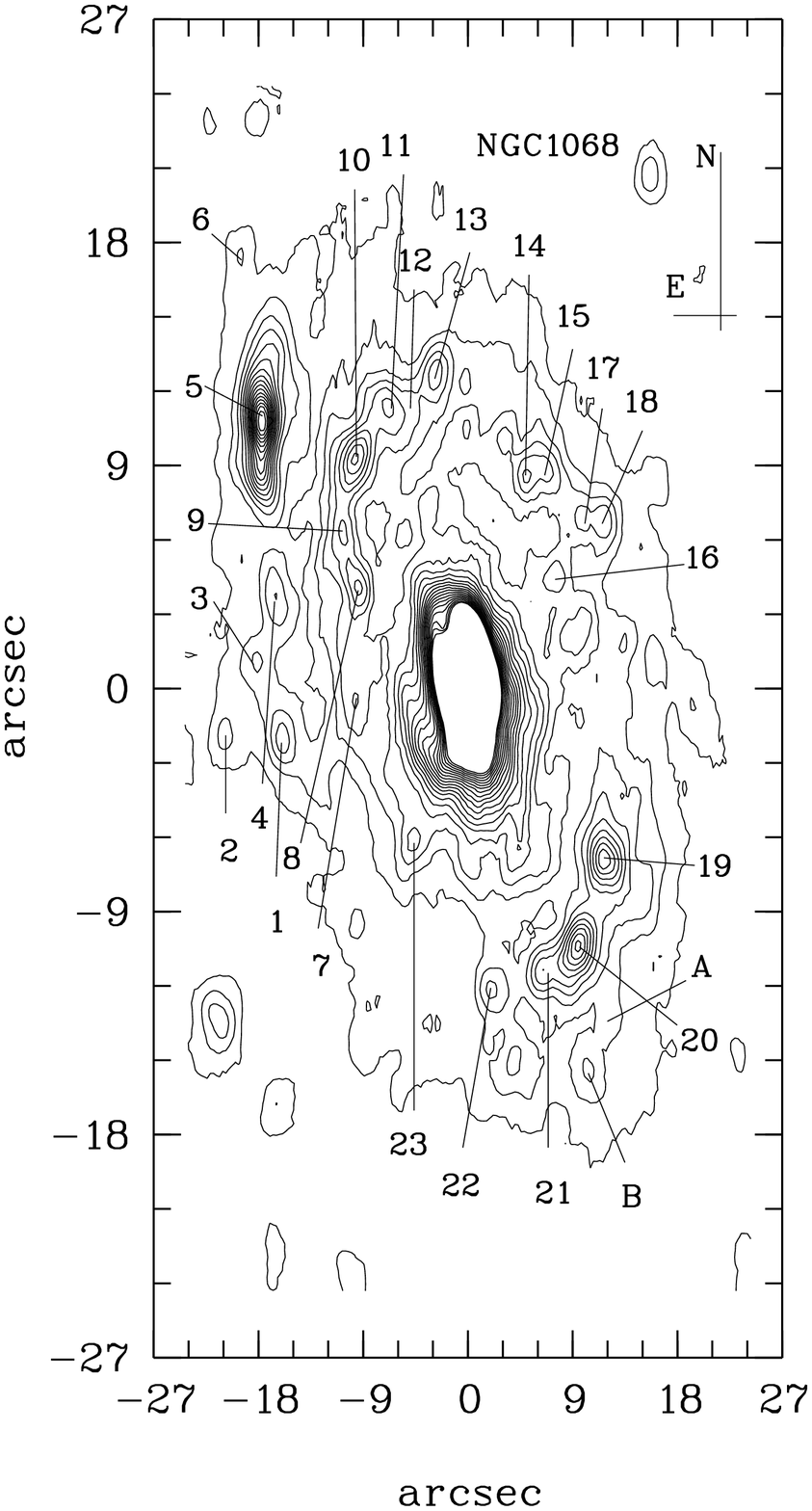,width=11cm,height=8.25cm}\nobreak
        \hspace{-25mm}
        \psfig{figure=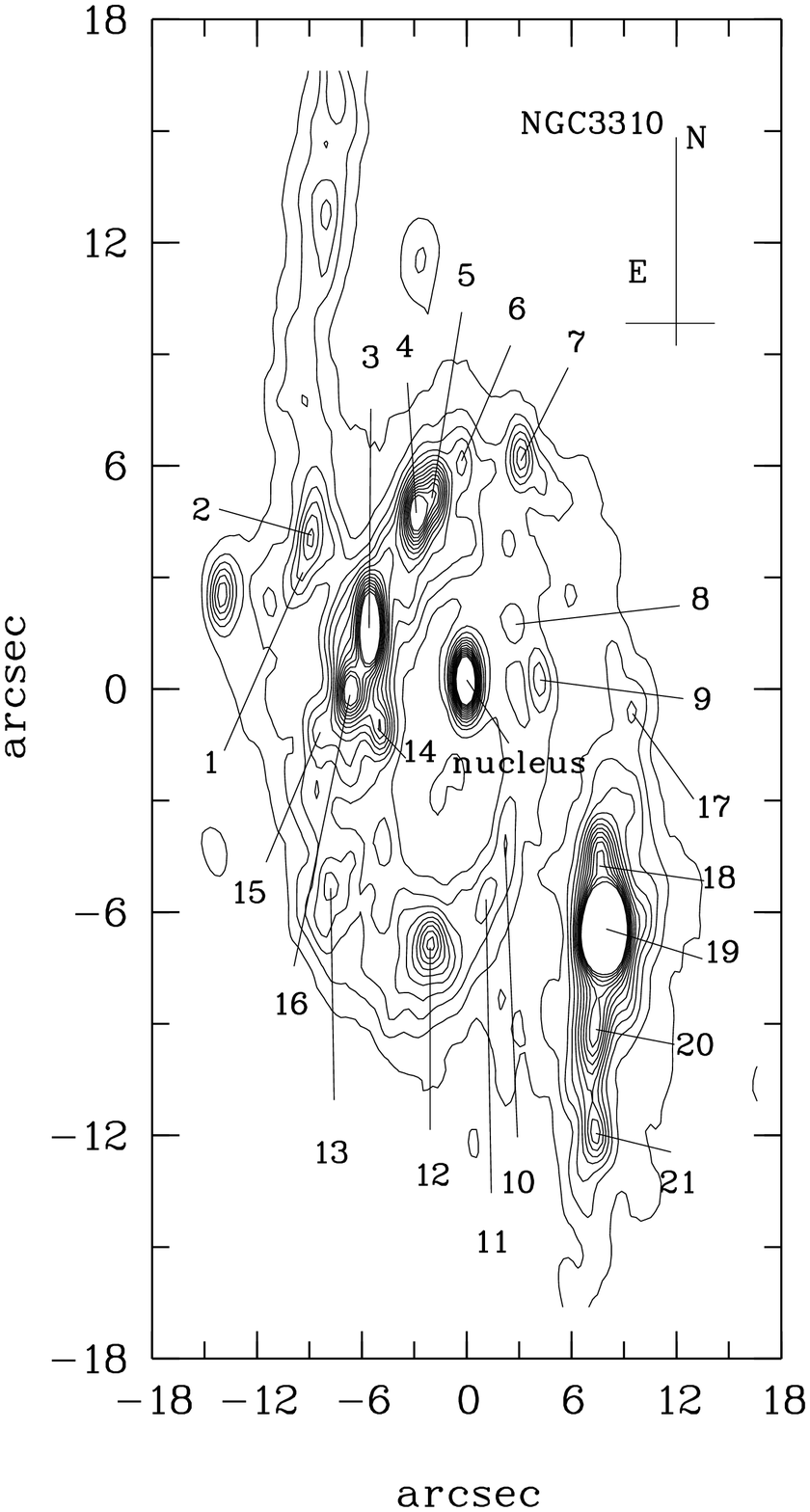,width=11cm,height=8.25cm}}
\vspace{12pt}    
  \hbox{\psfig{figure=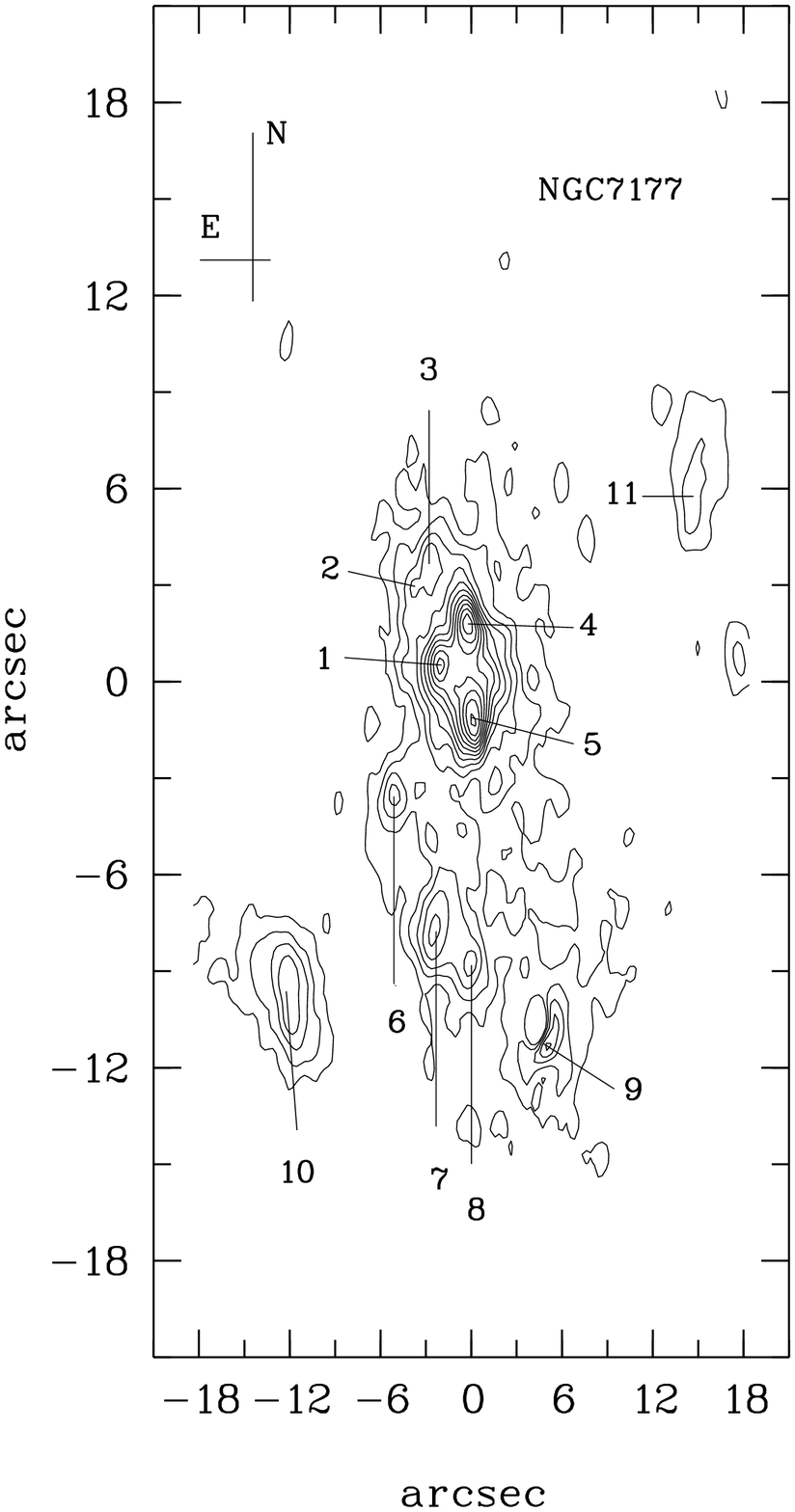,width=11cm,height=8.25cm}\nobreak
        \hspace{-25mm}
        \psfig{figure=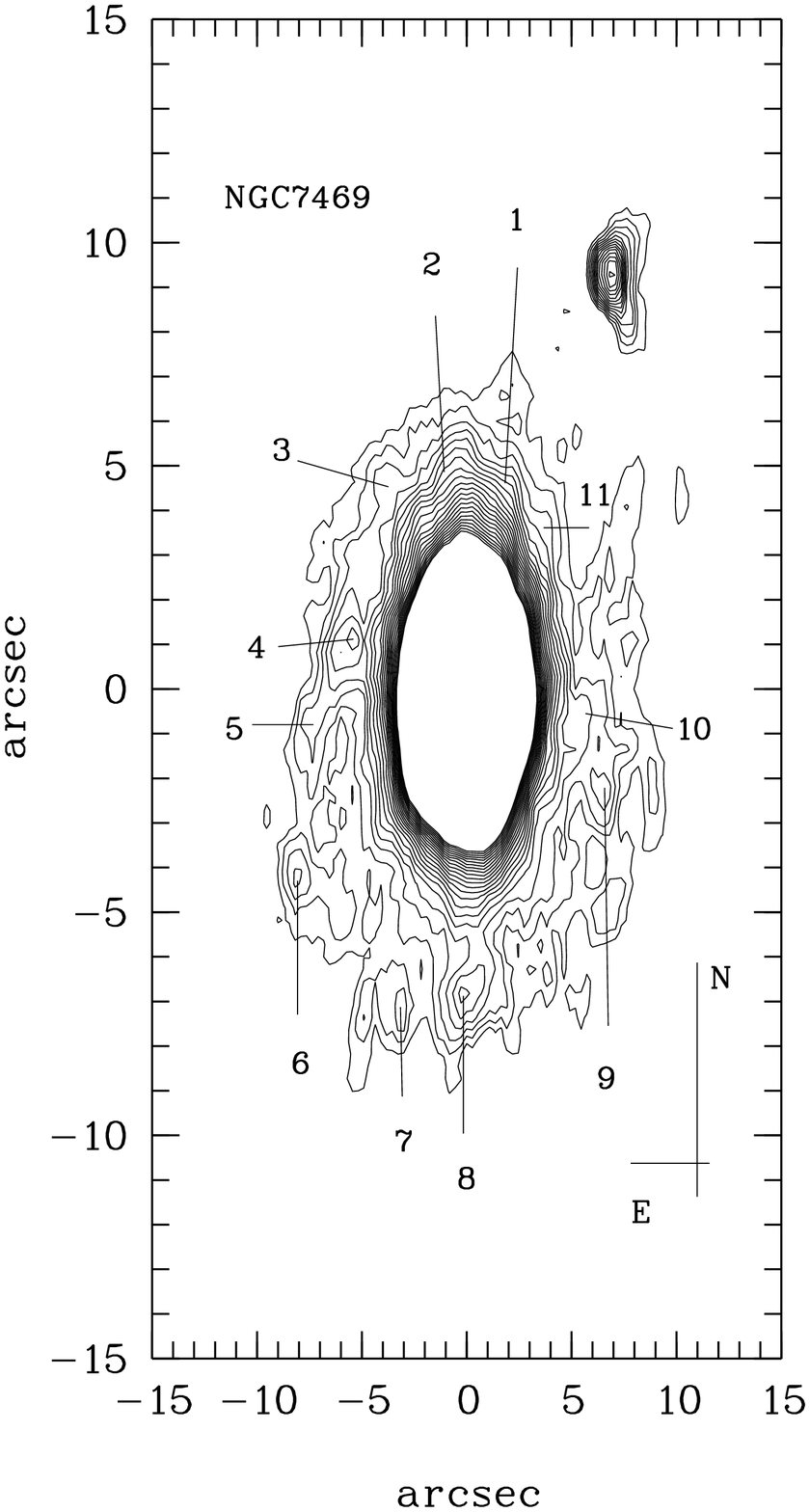,width=11cm,height=8.25cm}}
\caption{Isophotal maps of a) NGC~1068, b) NGC~3310, c) NGC~7177 and 
d) NGC~7469 with HII region identification labels} 
\end{minipage}
\end{figure*}

%
%
\begin{center}
\begin{figure*}
\begin{minipage}{150mm}
\psfig{figure=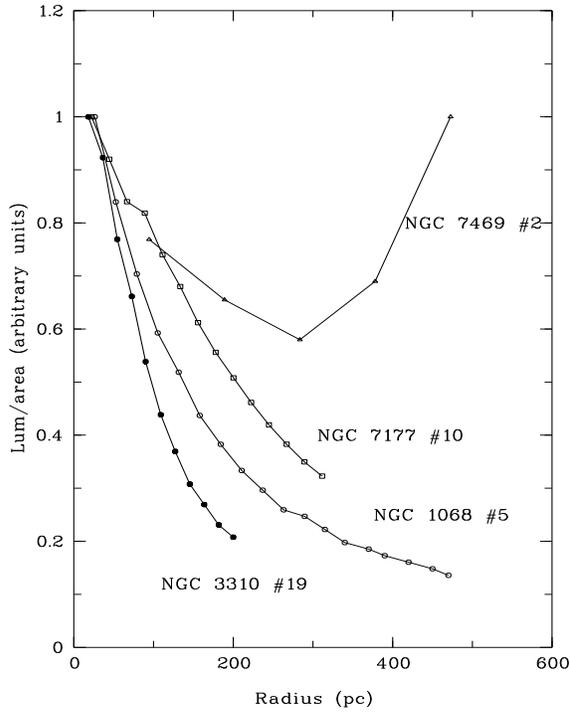,width=8cm,height=10cm}
\caption{Radial brightness profiles for selected HII regions. H$\alpha$ luminosity per unit area {\it vs} radius is plotted. The most regular behaviour 
corresponds to the biggest and most isolated regions of NGC~3310, NGC~1068 and
NGC~7177.  The asymptotic
radius is reached when the flux is about 10\% of the flux in
the central pixels of each region. The rise in the NGC~7469 region \# 2 profile
is due to the contribution of the nuclear flux.}
\end{minipage}
\end{figure*}
\end{center}


%
%

\begin{figure*}
\psfig{figure=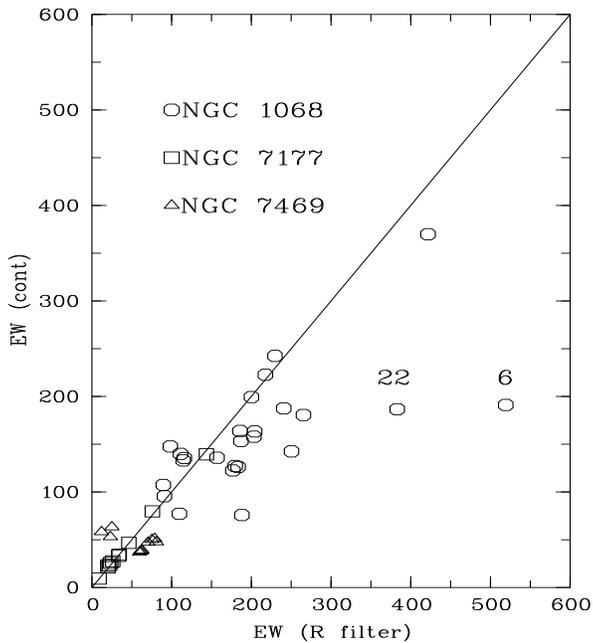,width=8cm,height=9cm}
\caption{Equivalent width of the line obtained using the continuum frame, 
{\it vs} that obtained using the R filter
for NGC1068, NGC7177 and NGC7469. The solid line stands for EW(cont) = EW(R filter). The scattering in the points is possibly due to insufficient background computation or slight missalignment.}
\end{figure*}


%

\begin{figure*}
\begin{minipage}{150mm}
\psfig{figure=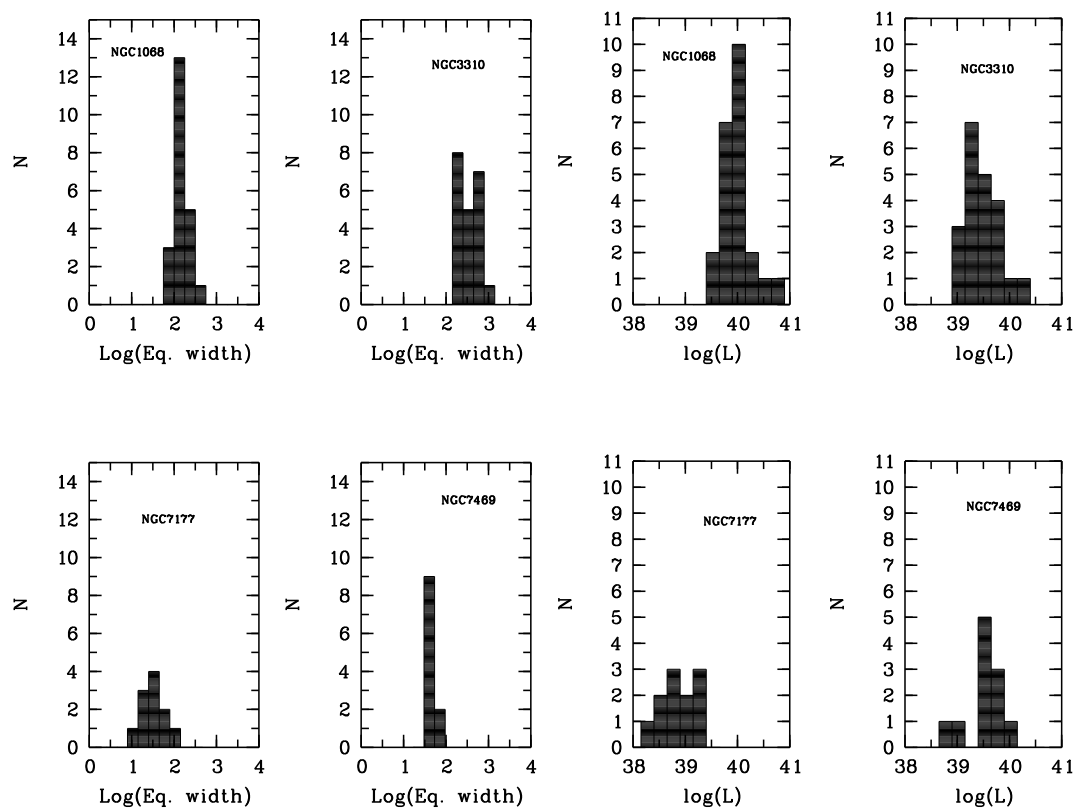,width=17cm,height=12cm,angle=270}
\caption{Histograms of luminosity and equivalent width of H$\alpha$ for NGC~1068, NGC~3310 and NGC~7469, the peak in H$\alpha$ luminosity is found at 
logL(H$\alpha$) = 39.5, while for NGC~7177 the luminosities seem to be, in
average, lower by an order of magnitude. The medium value of log(EW) is around 2.5 for NGC~1068 and NGC~3310, 1.9 for NGC~7469 and 1.5 for NGC~7177} 
\end{minipage}
\end{figure*}


%

\begin{figure*}
\begin{minipage}{150mm}
\psfig{figure=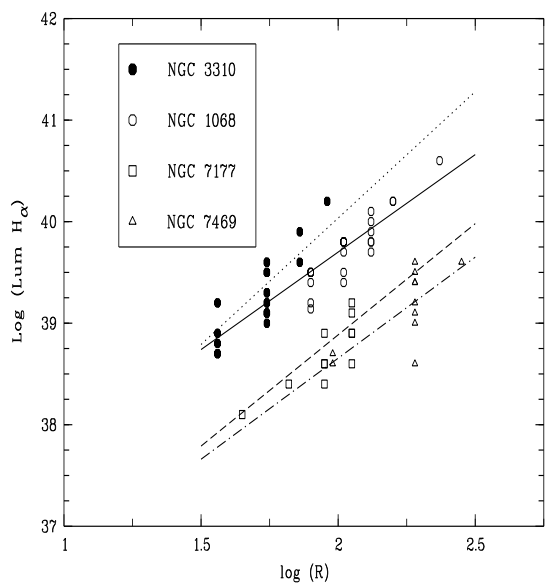,width=8cm,height=8cm,angle=270}
\caption{Logarithmic relation between H$\alpha$ luminosity radius for each individual HII region. The different lines represent functions log L = a log R + b where a is nearly 2 in all cases.}
\end{minipage}
\end{figure*}


%

\begin{figure*}
\begin{minipage}{150mm}
\psfig{figure=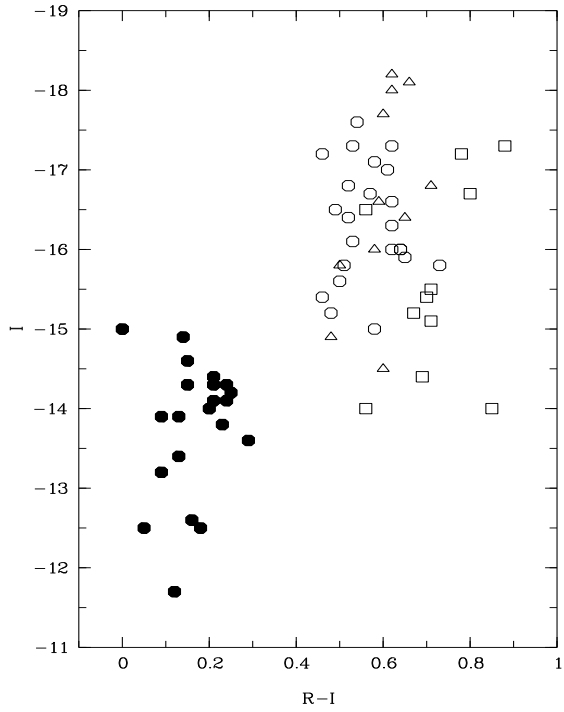,width=8cm,height=10cm}
\caption{ M$_I$ {\it vs} R-I colour-magnitude diagram for all the HII regions of the sample. Symbols as in Fig. 5}
\end{minipage}
\end{figure*}


%

\begin{figure*}
\begin{minipage}{150mm}
\psfig{figure=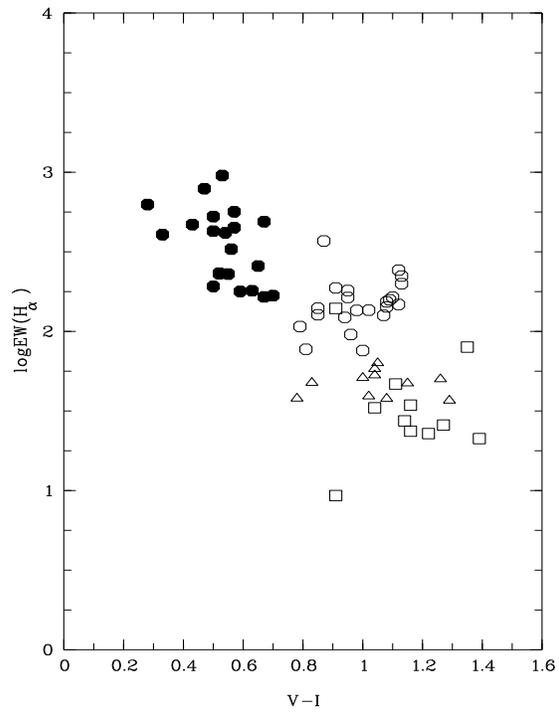,width=8cm,height=10cm}
\caption{ M$_I$ {\it vs} Equivalent width of H$\alpha$ {\it vs} V-I colour for all the HII regions of the sample. Symbols as in Fig. 5}
\end{minipage}
\end{figure*}

%

\begin{figure*}
\begin{minipage}{150mm}
\psfig{figure=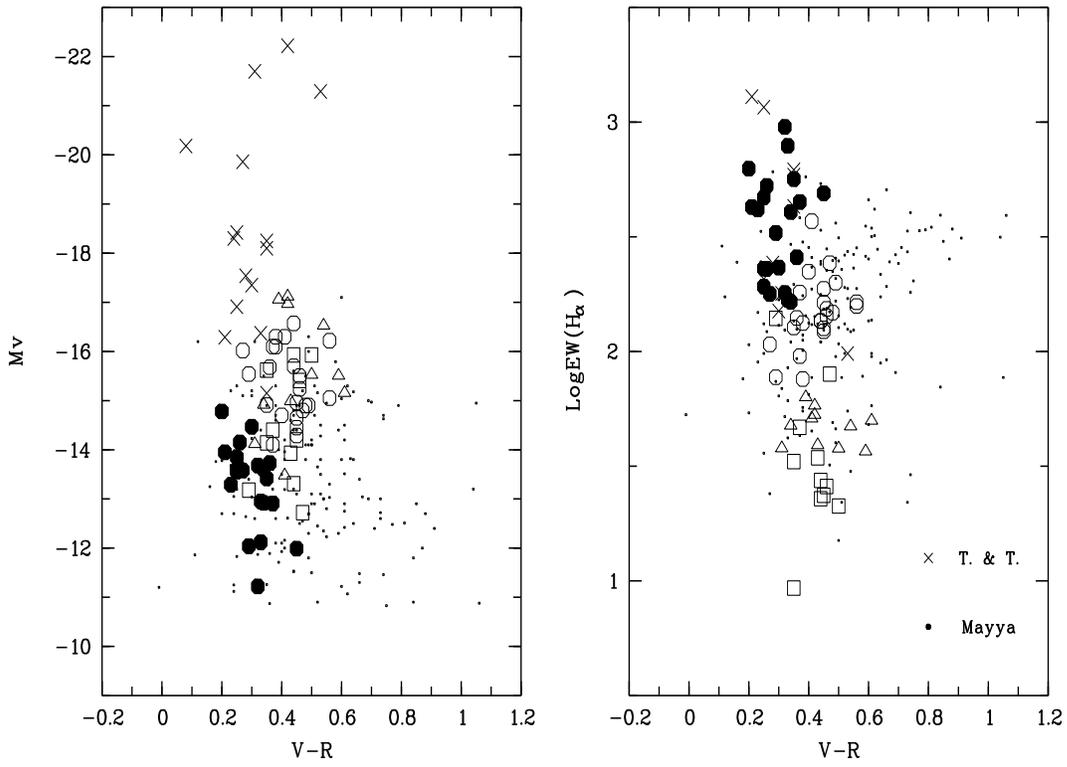,width=15cm,height=12cm,angle=270}
\caption{Colour-magnitude and colour {\it vs} logEW(H$\alpha$) for our data and
those found in the literature. Symbols as in Fig. 5 with added $\times$ from 
Telles \& Terlevich (1997) and dots from Mayya (1994).}
\end{minipage}
\end{figure*}


%
%
\begin{figure*}
\begin{minipage}{150mm}
\psfig{figure=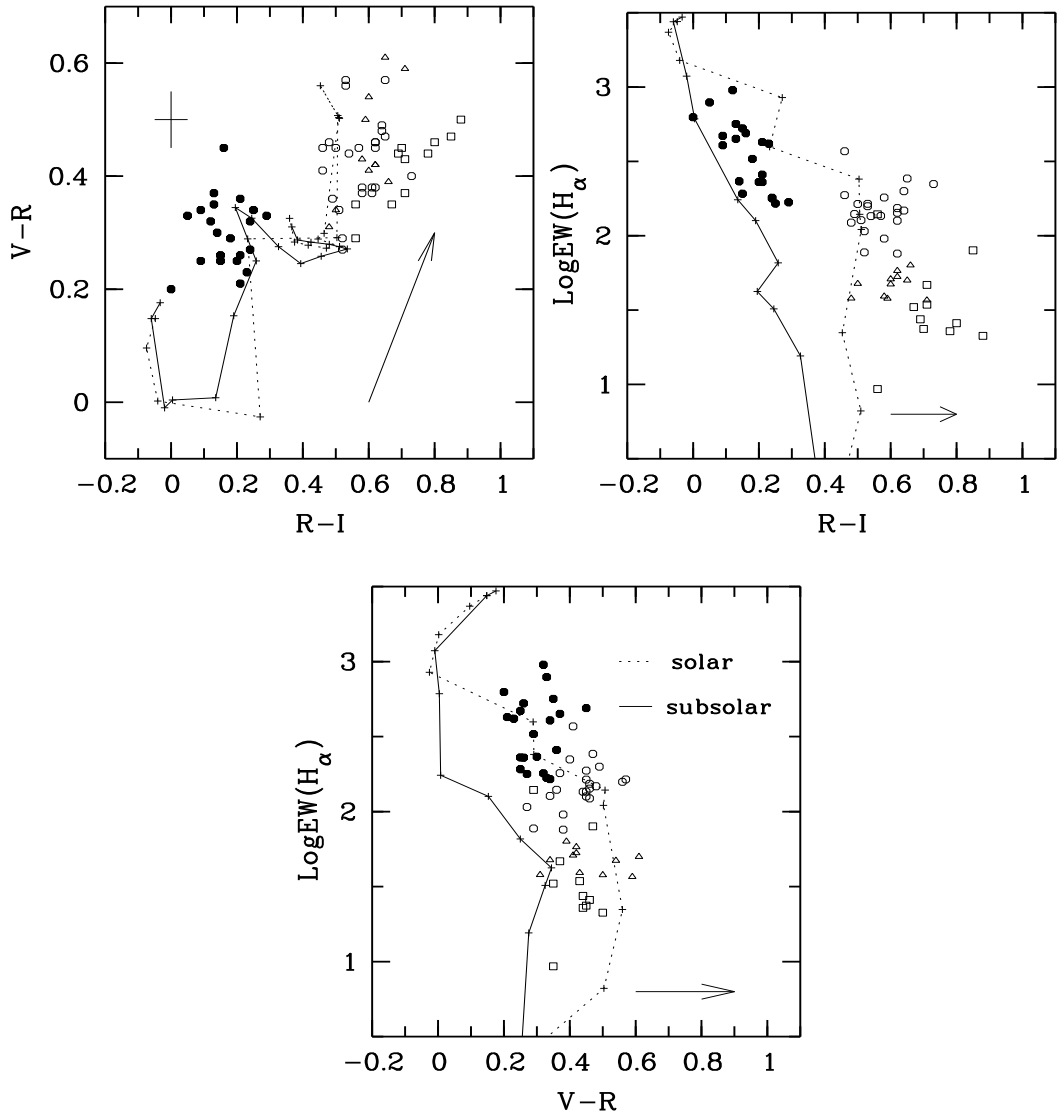,width=15cm,height=20cm}
\caption{V-R {\it vs} R-I and equivalent width of H$\alpha$ {\it vs} R-I and {\it vs} V-R diagrams for a single burst stellar population for solar and subsolar metallicities. The effect of reddening (Av=1mag) is plotted in the bottom right of the figures, as well as a typical error bar in the top-left one. Symbols as in Fig. 5}
\end{minipage}
\end{figure*}

%
%
\begin{figure*}
\begin{minipage}{150mm}
\psfig{figure=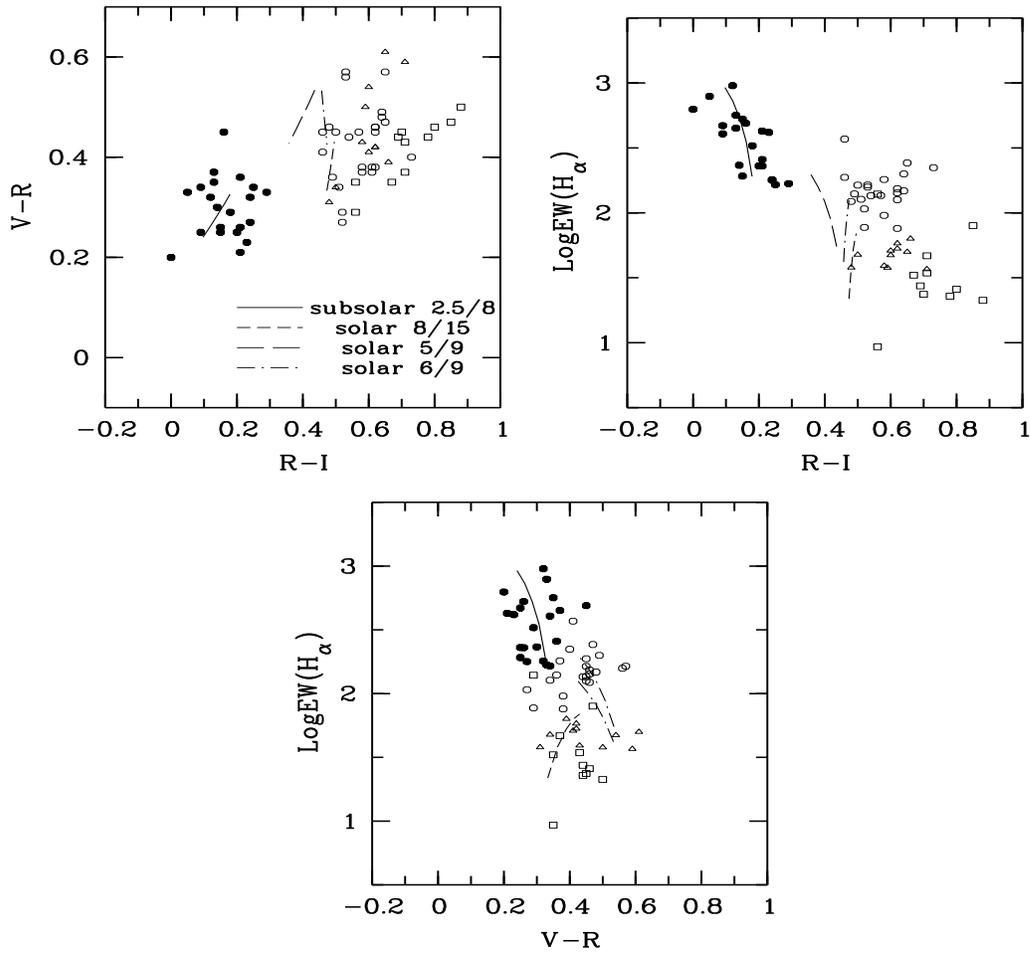,width=15cm,height=17cm}
\caption{Same as Fig. 9 but for a two-burst stellar population.}
\end{minipage}
\end{figure*}


%
%
\begin{figure*}
\begin{minipage}{150mm}
\psfig{figure=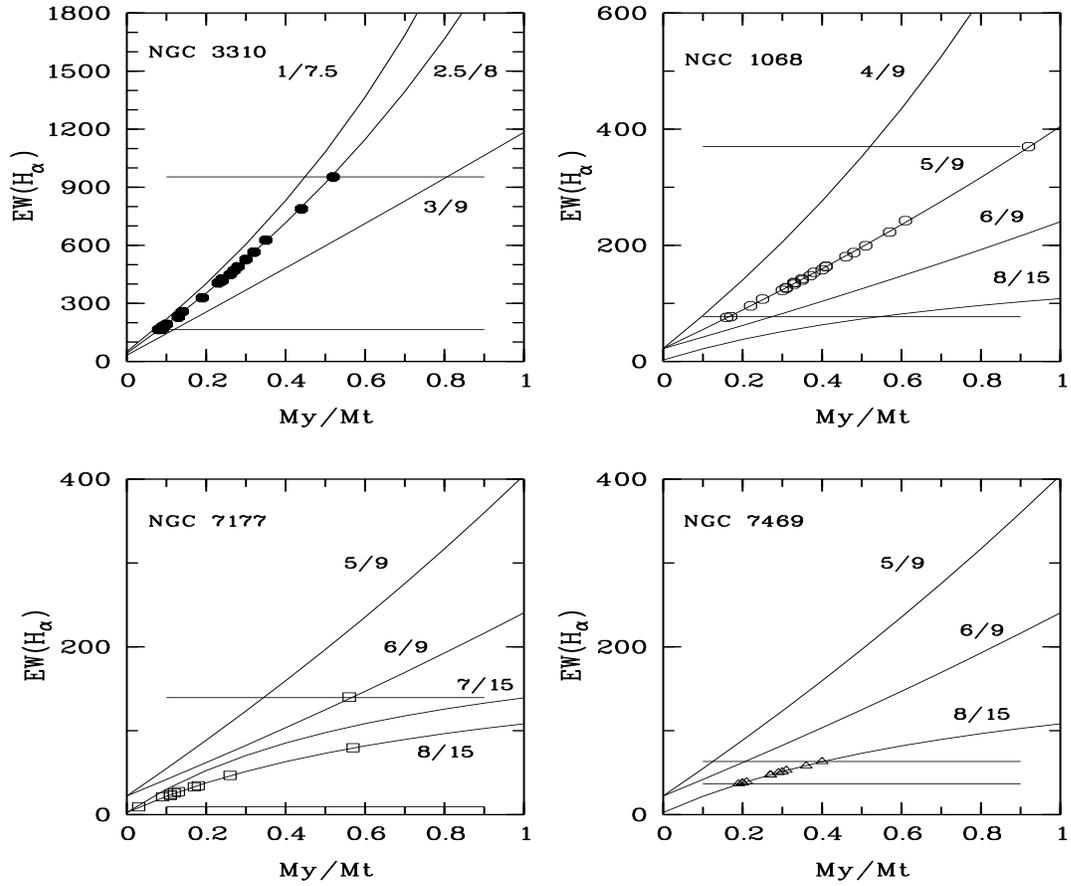,width=15cm,height=15cm}
\caption{Two-burst models that reproduce the equivalent width of H$\alpha$ , as a function of the fraction of the young burst mass. Horizontal lines 
correspond to the maximum and minimum values of EW(H$\alpha$) measured in each 
galaxy CNSFRs. Several options are displayed and the regions are overplotted over the chosen best couple of ages.}
\end{minipage}
\end{figure*}


\end{document}